\documentclass[fleqn,usenatbib]{mnras}

\usepackage{newtxtext,newtxmath}

\usepackage[T1]{fontenc}

\DeclareRobustCommand{\VAN}[3]{#2}
\let\VANthebibliography\thebibliography
\def\thebibliography{\DeclareRobustCommand{\VAN}[3]{##3}\VANthebibliography}

\usepackage{graphicx}	
\usepackage{amsmath,commath,mathtools}
\usepackage{float}
\usepackage{booktabs}
\usepackage{ulem}

\newcommand{\kvec}{\ensuremath{ {\boldsymbol k} }}

\usepackage{multirow}
\usepackage{color}

\definecolor{dullyellow}{rgb}{0.95, 0.8, 0.2}

\definecolor{darkgreen}{rgb}{0.0, 0.5, 0.0}

\title[Core-halo relations in SI-SFDM]{Core-halo scaling relations in self-interacting scalar field dark matter}

\author[Jessica N. López-Sánchez et al.]{
Jessica N. López-Sánchez,$^{1}$\thanks{E-mail: lopez@fzu.cz}
Erick Munive-Villa,$^{1}$\thanks{E-mail: munive@fzu.cz}
Tanja Rindler-Daller$^{2}$\thanks{E-mail: tanja.rindler-daller@univie.ac.at}
\\
$^{1}$CEICO—FZU, Institute of Physics of the Czech Academy of Sciences, Na Slovance 1999/2, 182 00 Prague, Czech Republic\\
$^{2}$Institut f\"ur Astrophysik, Universit\"atssternwarte Wien, Fakult\"at f\"ur Geowissenschaften, Geographie und Astronomie, \\
Vienna Int.School of Earth and Space Sciences, 
University of Vienna, T\"urkenschanzstr.17, 1180 Vienna, Austria \\
}

\date{Accepted XXX. Received YYY; in original form ZZZ}

\pubyear{\the\year{}}

\begin{document}
\label{firstpage}
\pagerange{\pageref{firstpage}--\pageref{lastpage}}
\maketitle

\begin{abstract}
We study the impact of self-interactions on the structure and evolution of scalar field dark matter (SFDM) halos. Using three-dimensional Gross–Pitaevskii–Poisson simulations of multiple soliton mergers, we explore both repulsive and attractive regimes across a wide range of scattering lengths. Our results show that repulsive self-interactions lead to more massive and extended cores with lower central densities compared to the free (non-interacting) fuzzy dark matter case, while attractive SI enhance central densities and can drive cores toward collapse, once a critical mass is exceeded. We confirm that the mass–radius relation of halo cores is well described by analytical predictions, even in the presence of self-interactions, and we extend the core–halo mass relation to scenarios beyond fuzzy dark matter. We find that the scaling relations between core mass, size, and total energy are not universal but depend sensitively on the strength and sign of the self-interaction, as well as on the evolutionary stage of the halo. These results demonstrate that self-interactions provide a natural mechanism to regulate core properties, with important implications for the formation of supermassive black holes and for potential astrophysical signatures in galactic cores.

\end{abstract}




\section{Introduction}
The fuzzy dark matter (FDM) model has gained popularity in recent decades, hosting the idea that dark matter (DM) can be described by a classical scalar field with ultralight masses around $m\sim 10^{-22}$ eV$/c^2$. 
This tiny value for the mass gives rise to one of the key features of this type of model, which makes it different from other DM candidates: the de Broglie wavelength, $\lambda_{\text{dB}}$, is of the order of kiloparsecs, encompassing the size of the galactic central regions \citep{lee1996galactic, hu2000fuzzy, Matos1999et, peebles2000fluid, matos2024short}. As a consequence, there is an effective quantum pressure that counteracts gravitational attraction and impacts the formation and evolution of structure. In fact, it is well known that FDM halos are characterized by a central core configuration, also known as a soliton, surrounded by a (properly averaged) Navarro-Frenk-White (NFW) envelope \citep{schive2014understanding}. This represents a stark contrast to the standard Cold Dark Matter (CDM) model, where halos tend to exhibit cuspy inner density profiles \citep{navarro1997universal}.

Regarding the phenomenology of the FDM model, different scaling relations concerning halo characteristics at large, compared to their central core properties, have been reported in the literature \citep{burkert2020fuzzy, nori2021scaling, chan2022diversity}. These relations provide a useful framework for making estimates of the galactic components and to compare with observations \citet{lanzoni2004scaling, araya2009cosmology, stanek2010massive, burkert2020fuzzy,lopez2024estimating}. 

In particular, the core-halo mass relation can provide insight into the interaction between the inner and outer regions of galactic systems. 
In the past, this relation has been inferred using mergers of idealized halos, cosmological simulations, or through semi-analytical approaches. However, the processes involved in halo formation and the impact on the core properties, which would allow us to estimate a universal scaling relation, are still subject of investigations. Understanding this phenomenon is crucial, since it provides a basis for comparison with observations which, at the same time, can be used to constrain the fundamental particle parameters of the FDM model. 

In \citet{schive2014understanding}, merger simulations of solitons in static and expanding backgrounds were performed to infer the following redshift-dependent core-halo mass relation $M_c\propto a^{-1/2}M^{1/3}$, where $a$ is the scale factor, and subscript $c$ denotes core properties. This relation has been derived by considering that halos are virialized systems and that the circular velocity at the halo virial radius is comparable to the circular velocity at the core radius or, in terms of the velocity dispersions (since they are of the same order of magnitude), $\sigma_c\sim\sigma_{h}$, see \citet{chavanis2019derivation}. On the other hand,  \citet{mocz2017galaxy} performed mergers of idealized solitons without cosmic expansion, characterizing the system by a different relation $M_c\propto M^{5/9}$. In that work, the authors proposed $M_c\sigma_c^2\sim M\sigma_{h}^2$. Additionally, in \citet{schwabe2016simulations}, a similar relation was reported, using major and minor binary and multiple soliton mergers. However, they also confirm that they could not reproduce the relation in \citet{schive2014understanding}. Furthermore, in \citet{du2017core}, the core-halo mass relation was deduced considering a semi-analytical approach for the history of halo mergers in terms of the mass loss fraction, $\beta$, in the core, leading to $M_c\propto M^{2\beta-1}$, finding that if $\beta=0.7$ they reproduce the results in \citet{schive2014understanding}. Then, in \citet{nori2021scaling}, a small number of halos were simulated in a cosmological framework, including the contributions of the environment and morphology. The authors claimed that it is possible to reproduce the results for both \citet{schive2014understanding} and \citet{mocz2017galaxy}, provided that relaxed systems are considered. They also mentioned that there should be a more generalized core-halo mass relation, which describes halos even if they are still in the process of relaxation. On the other hand, \cite{zagorac2023soliton} found that the scaling relation depends upon the simulation setup, and is not universal. Finally, in \citet{chan2022diversity}, mergers and cosmological simulations were performed, reporting a relation of the form $M_c=\beta+(M/\gamma)^\alpha$. The authors found that they can cover all previously found relations \citep{schive2014understanding, mocz2017galaxy, nori2021scaling}, for halos with sufficiently high velocity dispersion or mass, respectively. 

So far, the FDM model has been extensively explored to illustrate the diversity of scaling relations (such as the core-halo mass relation), depending on the complexity of the initial conditions, the type of simulation performed, and the population of halos taken into account. Through this process, several constraints have been imposed on the scalar field mass \citep{schutz2020subhalo, banik2021novel, nadler2021constraints, rogers2021strong, winch2024novel, sipple2025fuzzy}. For this and other reasons, the community has also studied alternative and more varied versions of the model by considering self-interactions (SI), for instance. In fact, if dark matter is an axion-like particle, then a second degree of freedom naturally arises in the theory, giving rise to attractive or repulsive self-interactions  \citep{2016JiJi, desjacques2018impact, chavanis2025review}. These DM particles may have different origins, including models motivated by open problems in QCD \citep{peccei1977cp, weinberg1978new} or string theories \citep{svrcek2006axions}. Depending on the model, the magnitude of the SI and the mass of the DM particles can vary, and the entire model family is often called scalar field dark matter (SFDM). Previous literature has shown that even in the presence of a weak SI, the impact on the cosmic structure exhibits significant differences with respect to the case without SI as follows. 

Recently, several authors have investigated the rich phenomenology of SFDM and its cosmic evolution in the linear and nonlinear regime. Beyond the effect of the quantum pressure given by the quantum nature of the scalar field, the self-interaction term introduces an additional force which can either counteract (repulsive interaction) or reinforce (attractive interaction) the gravitational force, depending on its sign. The former inhibits structure formation below the characteristic spatial scale given by the SI, while the latter enhances it. In fact, the evolution of matter depends on the strength of such an interaction, and this can be reflected in the abundance, distribution, and shape of SFDM halos.

In the regime of strongly repulsive SI, the so-called Thomas-Fermi (TF) regime of SFDM (with various names like "SFDM-TF", "SIBEC-DM", or "fluid dark matter"), this characteristic scale is close to the TF radius, which is the radius of a spherical, gravitational ($n=1$)-polytrope. This scale is much larger than the de Broglie scale for quantum pressure, yet both scales determine the halo properties. As a result, cosmological simulations are computationally even more expensive than in FDM, given the demands on resolution. Therefore, fluid approximations of the equations of motion, the Gross-Pitaevskii-Poisson equations, have been devised for this latter regime, first for 1D single-halo infall calculations in \citet{Dawoodbhoy2021beb, shapiro2022cosmological} and later extended to 3D within a cosmological framework in  \citet{2022Hartman,foidl2023halo}. 
The simulations in this regime revealed a core-envelope halo structure, similar to what was known from FDM simulations but with different characteristics. For instance, halo cores exhibit a larger size and lower density compared to the cores of FDM halos of the same total mass. A comparison of the fluid approach of SFDM with that of CDM allowed for further understanding of how the halo evolution proceeds over time. 
The CDM fluid picture follows naturally from the SFDM fluid equations, and reproduces very well the results from standard $N$-body simulations of CDM; see \citet{Dawoodbhoy2021beb, shapiro2022cosmological} and corresponding references therein. Furthermore, in \citet{foidl2023halo}, it was shown that the evolution of both SFDM-TF and CDM halos follows basically a two-stage process: in the early stage, cores are formed in both DM models. However, only SFDM-TF halos retain cored central parts throughout their evolution, because of the extra pressure support from repulsive SI. In contrast, CDM does not have this extra pressure, and halos will end up with cuspy density centers, akin to NFW profiles.

Apart from cosmological merger simulations for SFDM-TF using the fluid approach in \cite{2022Hartman}, along with a proof-of-principle merger simulation in \citet{foidl2023halo} (which was otherwise limited to single-halo infall simulations), there are barely any merger simulations for repulsive SFDM available, in general, let alone representative cosmological simulations. Among other things, binary soliton merger simulations with weakly to moderately repulsive SI have been performed in \citet{stallovits2024single}, showing a similar behaviour to that of the TF regime, namely that post-merger objects have larger cores with lower densities, compared to FDM.     
Furthermore, persistent oscillations in the envelopes of post-merger halos have been found, in accordance with previous FDM simulations, but these oscillations have a smaller amplitude, once a repulsive SI is added. Similar results have been reported in the work by \citet{galazo2024solitons}, which focused on the TF regime, and further differences between FDM and models with (moderate) SI have been found in \cite{indjin2025fuzzy}.

The results of this extensive body of literature, and further references therein, have determined that SFDM models, even with small values of repulsive SI, can show marked differences to FDM.

On the other hand, SFDM with attractive SI has received more attention in previous literature. In this regime, there is also a critical mass scale for DM halos which gives rise to two different possibilities in their evolution: the central part of SFDM halos may remain a dilute soliton if the SI is weak, or it may become a dense soliton, which is unstable to the possible formation of supermassive black holes (SMBH) \citep{chavanis2011mass, Avilez_etal2018, padilla2021core, mocz2023cosmological, painter2024attractive}, for sufficiently high values of the attractive SI. 


Since the evolution of the halos changes for different SI regimes, it is expected that this physics also has an impact on various scaling relations, particularly the core-halo mass relation described above. To the best of our knowledge, this question has not yet been investigated at the level of numerical simulations, although analytical predictions for various cases have been presented in \citet{padilla2021core}.   

In this paper, our aim is to explore various core-halo scaling relations, notably the core-halo mass relation and some core properties in particular, for SFDM in several SI regimes.  We perform merger simulations of up to 50 solitons within the framework of solving the Gross-Pitaevskii wave equation of motion, coupled to the Poisson equation. In particular, we include the study of the TF regime of strongly repulsive SI without resorting to fluid approximations. This is feasible because we consider a small, static simulation box for all the runs that we perform. Apart from the TF regime, we include models with weak to high values of attractive SI, as well as models without SI, for comparison.

The paper is organized as follows. In Section \ref{sec: numerics}, we describe the numerical implementation used to solve the Gross-Pitaevskii-Poisson system of differential equations. In Section \ref{sec: simulations}, the simulation setup is described as well as the density profiles of the final halo configurations of each SFDM model. A detailed analysis of various scaling relations, pertaining to the core, as well as important core-halo relations, for different SI regimes and their evolution over time, is presented in Section \ref{sec: core}.
In Section \ref{sec: self-similar}, we explore the accretion of mass in the core for different scenarios of the SFDM model, using self-similar solutions. Finally, in Section \ref{sec: conclusions}, we sketch some conclusions, regarding the phenomenology of the attractive and repulsive interactions of the SFDM model.

\section{Self-interacting SFDM in the non-relativistic regime}\label{sec: numerics}

\subsection{The Gross-Pitaevskii-Poisson equations}\label{sec: GPP}
The non-relativistic Gross-Pitaevskii-Poisson (GPP) equations describe the evolution of the wave function $\psi$ of the SFDM condensate with self-gravity:
\begin{eqnarray}\label{GPP}
    i\hbar\frac{\partial\psi}{\partial t}&=&-\frac{\hbar^2}{2m}\nabla^2\psi+m\Phi\psi+\frac{4\pi \hbar^2\rho_{0} a_s}{m^2}\abs{\psi}^2\psi \nonumber \\
     &&+\frac{32\pi \hbar^4\rho_{0}^2\abs{a_s}^2}{3m^5c^2}\abs{\psi}^4\psi \nonumber \\
    \nabla^2\Phi &=& 4\pi G \rho_{0}(\abs{\psi}^2-1),
\end{eqnarray}
where $\hbar$ is the reduced Planck constant, $G$ is the gravitational constant, and $m$ is the mass of the SFDM particle. In this paper, we assume spin-$0$ bosons. The squared amplitude of the wave function is related to the density. As is customary in structure-formation simulations, it is normalized such that 
$\abs{\psi}^2=\displaystyle\frac{\rho}{{\rho_0}}$ with $\rho_0$ the mean density within the simulation box. 
Furthermore, the GP equation includes self-interaction (SI) terms which model short-range s-wave scattering between two bosons up to second order, where we will choose a range of values for the scattering length $a_s$. The first-order term greatly dominates the scattering dynamics, thus values of $a_s<0$ or $a_s>0$ indicate overall attractive or repulsive SI, respectively. SFDM models without self-interaction, i.e. FDM, have $a_s=0$; in this case, the equations have also been known in the literature as the Schr\"odinger-Poisson system. 

We have included the higher-order term proportional to $\psi^5$ in the Gross–Pitaevskii equation \eqref{GPP}, as also discussed in \citet{, chavanis2016collapse, mocz2023cosmological, painter2024attractive}. In Appendix~\ref{app: comparison}, we compare the evolution of systems with and without this higher-order contribution, focusing on the density evolution and energy conservation results, which are of relevance in the following sections. The figures in Appendix~\ref{app: comparison} exemplify that the impact of the higher-order contribution is very small.


The prefactor in the cubic SI term includes the self-coupling strength $g=\displaystyle4\pi\hbar^2\frac{a_s}{m}$.
In what follows, we will consider a dimensionless form of the self-coupling, and for that we use the expression
\begin{equation}
\Lambda=\displaystyle\frac{c^2 g}{4\pi G\hbar^2},
\end{equation}
see e.g. \citet{padilla2021core} (it basically stems from a proper reduction of the relativistic to the non-relativistic limit of interest to us here). The parameter $\Lambda$ is useful to distinguish between the strong and weak regimes of self-interactions. If $\Lambda \gg 17$, SFDM is in the TF regime, and solitons, resp. halo cores, are governed by a balance between gravity and repulsive SI, see e.g. \citet{Rindler-Daller:2011afd,Dawoodbhoy2021beb} for more details. 

\subsection{Ground-state solutions}

The simulations are initialized with stationary soliton configurations.
In general, stationary equilibrium configurations can be obtained by expressing the wave function as
\begin{equation}
    \psi = e^{-i\omega t/\hbar} \psi_{\text{sol}}(r)
\end{equation}

\noindent leading to the following spherically symmetric GPP system 
\begin{eqnarray}\label{eq: GPP_spherical}
    -\frac{\hbar^2}{2m}\frac{1}{r}\frac{d^2}{d r^2}(r\psi_{\text{sol}})&=&\bigg( \omega  -m \Phi - \frac{4\pi\hbar^2\rho_0a_s}{m^2}\psi_{\text{sol}}^2 \nonumber \\
  &&  -\frac{32\pi\hbar^4\rho_0^2\abs{a_s}^2}{3m^5c^2}\psi_{\text{sol}}^4 \bigg)\psi_{\text{sol}}\\
    \frac{1}{r}\frac{d^2}{d r^2}(r\Phi)&=& 4\pi G\rho_0\psi_{\text{sol}}^2.
\end{eqnarray}

It is convenient to rewrite the system of equations using the transformations $\hat{\psi}_{\text{sol}}=\displaystyle\frac{m\sqrt{G\rho_0}}{\hbar}\psi_{\text{sol}}$ and $\hat{\Phi}=\displaystyle\frac{m^2}{\hbar^2}\Phi$. Then, the above equations take the form
\begin{eqnarray}\label{eq: eigensystem}
    \frac{1}{2}\frac{d^2}{d r^2}(r\hat{\psi}_{\text{sol}})&=&r\hat{\psi}_{\text{sol}}\bigg(\hat{\Phi} - \hat{\omega} + \hat{a}_s \hat{\psi}_{\text{sol}}^2 \nonumber\\
    && +\frac{2}{3\pi}\lambda_{\text{compton}}^2\abs{\hat{a}_s }^2\hat{\psi}_{\text{sol}}^4\bigg)\\
    \frac{d^2}{d r^2}(r\hat{\Phi})&=&4\pi r \hat{\psi}_{\text{sol}}^2,
\end{eqnarray}
where $\hat{\omega}=\displaystyle\frac{m \omega}{\hbar^2}$ is a constant which corresponds to the eigenvalue of the system, $\hat{a}_s = \displaystyle\frac{4\pi a_s \hbar^2}{m^3 G} $ and $\lambda_{\text{compton}}=\displaystyle\frac{\hbar}{mc}$. This way, we make sure that the following conditions are met \citep{guzman2004evolution}, 
\begin{align}\label{eq: scaling_relations}
    \begin{aligned}
        &\hat{\psi}_{\text{sol}}(r\rightarrow \infty) \rightarrow 0, & \quad &\hat{\Phi}(r\rightarrow\infty) = -\frac{GM m^2}{\hbar^2r}, \\
        &\hat{\psi}(r\rightarrow 0) = 1, & \quad &\frac{\partial\hat{\Phi}}{\partial r}\bigg|_0 \rightarrow 0, \\
        &\frac{\partial\hat{\psi}}{\partial r}\bigg|_0 \rightarrow 0, & \quad &\frac{\partial\hat{\psi}}{\partial r}\bigg|_{r\rightarrow\infty} \rightarrow 0,
    \end{aligned}
\end{align}
where $M(r)=\int \rho dV=4\pi\int \rho(r)r^2dr$ is the enclosed mass at radius $r$, and $M$ denotes the total mass. Given $\hat{\Phi}$ and $a_s$, the system of equations is reduced to an eigenvalue problem with unique values for $\omega$ and 
$\hat{\Phi}(0)$ 

for which such boundary conditions are fulfilled. In general, there is a discrete number of eigenvalues $\{\omega_0, \omega_1, \ldots, \omega_N\}$ associated with the number of nodes $N$ in the solution of $\hat{\psi}$. Here, however, we will focus on ground-state solutions without nodes, i.e. 
$\omega$ corresponds to $\omega_0$. 

The GPP system is invariant under the following transformations
\begin{equation}\label{eq: scaling1}
    \left\{t, x , \psi, \rho, a_s\right\} \to \left\{\lambda^{-2}\hat{t}, \lambda^{-1}\hat{x} , \lambda^{2}\hat{\psi},\lambda^{4}\hat{\rho}, \lambda^{-2}\hat{a}_s\right\},
\end{equation}
and therefore,
\begin{equation}
 \left\{M, E \right\} \to \left\{\lambda\hat{M}, \lambda^{3}\hat{E} \right\}. 	
\end{equation}

Furthermore, the GPP system also scales under transformation of the mass of the scalar boson $m\rightarrow \alpha \hat{m}$ as follows,

\begin{equation}\label{eq: scaling2}
    \left\{t, x , \psi, \rho, a_s\right\} \to \left\{\alpha\hat{t}, \hat{x} , \alpha^{-1}\hat{\psi},\alpha^{-2}\hat{\rho}, \alpha^{3}\hat{a_s}\right\},
\end{equation}
and therefore,
\begin{equation}
 \left\{M, E \right\} \to \left\{\alpha^{-2}\hat{M}, \alpha^{-4}\hat{E} \right\}. 	
\end{equation}

Stability tests for soliton solutions with SI are shown in Appendix
\ref{app: stability}, where we can see that energy conservation is well established in the simulations.

\subsection{Numerical implementation}
\label{numerical_implementation}

There are different numerical approaches to solving the GPP system \eqref{GPP} reported in the literature; see e.g. \citet{Edwards_2018, Guzman2018evm, mocz2018schrodinger, zhang2018ultralight, munive2022solving}. Here, we will make use of the pseudo-spectral method by modifying the code developed in \citet{lopez2025scaling}. It uses a pseudo-spectral time-splitting method optimized with graphics processing units (GPUs) using CUDA. This code was used to study general spin-$s$ SFDM models. Here, we will focus on the spin-$0$ case, but we now implement SI terms in the equation of motion. This can be done by rewriting the GP equation as 
\begin{equation}\label{eq: new_schrodinger}
    i \hbar \frac{\partial \psi}{\partial t} = -\frac{\hbar^2}{2m}\nabla^2\psi+\tilde{\Phi}\psi 
\end{equation}
where 
\begin{equation} \label{eq:fullpot}
\tilde{\Phi}=m\Phi+\displaystyle\frac{4 \pi \hbar^{2}\rho_0a_{s}}{m^{2}}|\psi|^{2}+ \frac{32\pi \hbar^4\rho_{0}^2\abs{a_s}^2}{3m^5c^2}\abs{\psi}^4,
\end{equation}
can be treated as an effective potential which encapsulates the nonlinear contributions of SI. This approach was also used in \citet{Dawoodbhoy2021beb, glennon2021modifying, stallovits2024single}.

Note that eq. \eqref{eq: new_schrodinger} recovers the form of the Schr\"odinger equation which describes the FDM case (without self-interactions); then, we can apply the kick-drift-kick method described in \citet{lopez2025scaling}. In this scheme, the wave function and the potential are updated using the time-split scheme which is $\mathcal{O}$(2) in time. The evolution is described by the equations

\begin{subequations}
\begin{align}\label{eq: kdk:a}
    \mathbf{K} (\Delta t/2):\; & \psi(t) \to \text{e}^{-\frac{i m \Delta t \tilde{\Phi}(t)}{2 \hbar}} \psi(t),\\\label{eq: kdk:b}
    \mathbf{D} (\Delta t):\;&\psi(t+\Delta t/2) \to \mathcal{F}^{-1}\left[
    \text{e}^{- \frac{ i \Delta t \hbar \kvec^2}{2 m }}
    \mathcal{F} \psi(t+\Delta t/2) \right], \\\label{eq: kdk:c}
    \mathbf{K} (\Delta t/2):\;&\psi(t+\Delta t) \to \text{e}^{-\frac{i \Delta t m \tilde{\Phi}(t + \Delta t)}{2 \hbar}} \psi(t+\Delta t/2).
\end{align}
\label{eq: kdk}
\end{subequations}
$\mathcal{F}$ and $\mathcal{F}^{-1}$ stand for the discrete Fourier transformation and its inverse, respectively, $\mathbf{K}$ and $\mathbf{D}$ stand for \textit{Kick} and \textit{Drift}, respectively. Note that for the last Kick in equation \eqref{eq: kdk:c} we are expected to know the value of $\psi(t+\Delta t)$ in advance. However, this step is just a change of phase, which does not affect $|\psi|^2$, see also \cite{Edwards_2018}. Additionally, for attractive SI, we implemented a more stable time evolution given by an $\mathcal{O}(4)$ KDK method (see \cite{Tidal_O6}). The approach can be described simply as 
\begin{equation}
\begin{aligned}
\psi(t+\Delta t) =\;
&\mathbf{K}(v_2 \Delta t)\, 
 \mathbf{D}(t_2 \Delta t)\,
 \mathbf{K}(v_1 \Delta t)\, 
 \mathbf{D}(t_1 \Delta t)\, \\
&\mathbf{K}(v_0 \Delta t)\,
 \mathbf{D}(t_1 \Delta t)\,
 \mathbf{K}(v_1 \Delta t)\,
 \mathbf{D}(t_2 \Delta t)\,
 \mathbf{K}(v_2 \Delta t)\, \psi(t)
\end{aligned}
\end{equation}
where,
\begin{equation}
\begin{array}{l}
v_{1}=\displaystyle\frac{121}{3924}(12-\sqrt{471}), \quad w=\sqrt{3-12 v_{1}+9 v_{1}^{2}}, \\
t_{2}=\displaystyle\frac{1}{4}\left(1-\sqrt{\frac{9 v_{1}-4+2 w}{3 v_{1}}}\right), \quad t_{1}=\frac{1}{2}-t_{2}, \\
v_{2}=\displaystyle\frac{1}{6}-4 v_{1} t_{1}^{2}, \quad v_{0}=1-2\left(v_{1}+v_{2}\right).
\end{array}
\end{equation}
Note that after every $\mathbf{D}\mathbf{K}$ application, we need to solve the Poisson equation:

\begin{equation}
    \Phi = \mathcal{F}^{-1}\left[-\frac{1 }{\kvec^2} \mathcal{F} \left( 4\pi G \rho_0 \left( \abs{\psi}^2-1\right)\right) \right],
\end{equation}
where $\kvec$ is the wave vector in the frequency domain. Note that we follow a similar reasoning as before, and $|\psi|$ does not depend on phase shifts.

\section{Simulations of multiple soliton mergers}\label{sec: simulations}

In this paper, we perform simulations in three different SI regimes, as follows. The \textit{fuzzy regime} corresponds to models in which SI can be neglected. In this case, hydrostatic equilibrium is given by the balance between quantum pressure and gravity. The opposite regime is the \textit{Thomas-Fermi regime}, when hydrostatic equilibrium is due to gravity and repulsive SI. On the other hand, there is also a \textit{strongly attractive regime}, such that hydrostatic equilibrium gets close to instability with respect to gravitational collapse. The values for the scattering length $a_s$ considered in this work are summarised in Table \ref{tab: table_params}, where we also include the corresponding numbers for the dimensionless coupling $\Lambda$. In the following sections, scaling relations will be plotted for a selected subset of the parameters listed in Table \ref{tab: table_params}, in order to make the visualizations more legible.

We stress that for higher values of attractive SI ($|a_s| > 1\times 10^{-77}$ cm), the code becomes unstable and the evolution of the system stalls. For that reason, we focus on four attractive cases, with the highest value of $|a_s| = 6.62\times 10^{-77}$ cm, which remains stable up to a certain time before the central density diverges (see Appendix \ref{app: stability}).

Our work in this paper demonstrates that the pseudo-spectral method, described in Section \ref{numerical_implementation}, can be applied to simulations of SFDM models with strongly repulsive SI, i.e. for SFDM-TF models which have been previously simulated, using fluid approximations. The analytical treatment of the nonlinear term in Fourier space reduces the numerical errors that commonly arise in finite-difference schemes. In practice, SI contributes only a local phase factor at each step, which the algorithm can handle without introducing instabilities. However, the main requirements are a sufficiently high spatial resolution, in order to capture the sharp small-scale density gradients that emerge at strong repulsion, and to use a timestep fine enough to resolve the rapid nonlinear phase evolution. When these conditions are met, the pseudo-spectral approach remains both stable and accurate over a broad range of interaction strengths.

\begin{table}
    \centering
    \begin{tabular}{c|c|c}
        \textbf{Model} & $\Lambda$ & $a_s$ (cm) \\
        \hline
        \multirow{10}{*}{Repulsive} &           $3.0\times 10^7$ & $3.96\times 10^{-76}$ \\
         & $1.0\times 10^7$ & $1.32\times 10^{-76}$ \\
         & $6.0\times 10^6$ & $7.94\times 10^{-77}$ \\
         & $5.0\times 10^6$ & $6.62\times 10^{-77}$ \\
         & $2.0\times 10^6$ & $2.65\times 10^{-77}$ \\
         & $1.0\times 10^6$ & $1.32\times 10^{-77}$ \\
         & $4.8\times 10^5$ & $6.35\times 10^{-78}$ \\
         & $4.3\times 10^5$ & $5.69\times 10^{-78}$ \\
         & $3.6\times 10^5$ & $4.76\times 10^{-78}$ \\
         & $2.0\times 10^5$ & $2.65\times 10^{-78}$ \\
         & $1.1\times 10^5$ & $1.45\times 10^{-78}$ \\
         & $1.0\times 10^5$ & $1.32\times 10^{-78}$ \\
        \hline
        Free / no SI & 0.0 & 0.0 \\ 
        \hline
        \multirow{3}{*}{Attractive} & $-1\times 10^3$ & $-1.32\times 10^{-80}$ \\
        & $-1\times 10^4$ & $-1.32\times 10^{-79}$ \\
        & $-1\times 10^5$ & $-1.32\times 10^{-78}$ \\
         & $-5\times 10^6$ & $-6.62\times 10^{-77}$ \\
        \hline
    \end{tabular}
    \caption{Values of the dimensionless self-coupling term $\Lambda$ and the corresponding scattering length $a_s$ in cm, for the repulsive and the attractive cases considered in this work.}
    \label{tab: table_params}
\end{table}

\subsection{Numerical setup}

We performed simulations of multiple soliton mergers by varying the number of initial solitons $N_{\text{sol}}=\begin{pmatrix}
  10, & 15, & 20, & 25, & 30, & 35, & 40, & 45, & 50 
\end{pmatrix}$. The box size used was $L=100$ kpc with a mesh grid of $512^3$. The mass of the SFDM particle is fixed to a value of $m=10^{-22} ~\text{eV}/c^2$. All simulations were evolved until $t = 10 \tau_{\text{dyn}}$, where $\tau_{\text{dyn}} = \displaystyle\frac{1}{\sqrt{G\rho_0}}$ and run with periodic boundary conditions. When we repeated them using an artificial sponge, the results remained unchanged, showing that the sponge is not needed in this context.


%

\begin{figure*}
    

    \includegraphics[width=0.9\textwidth]{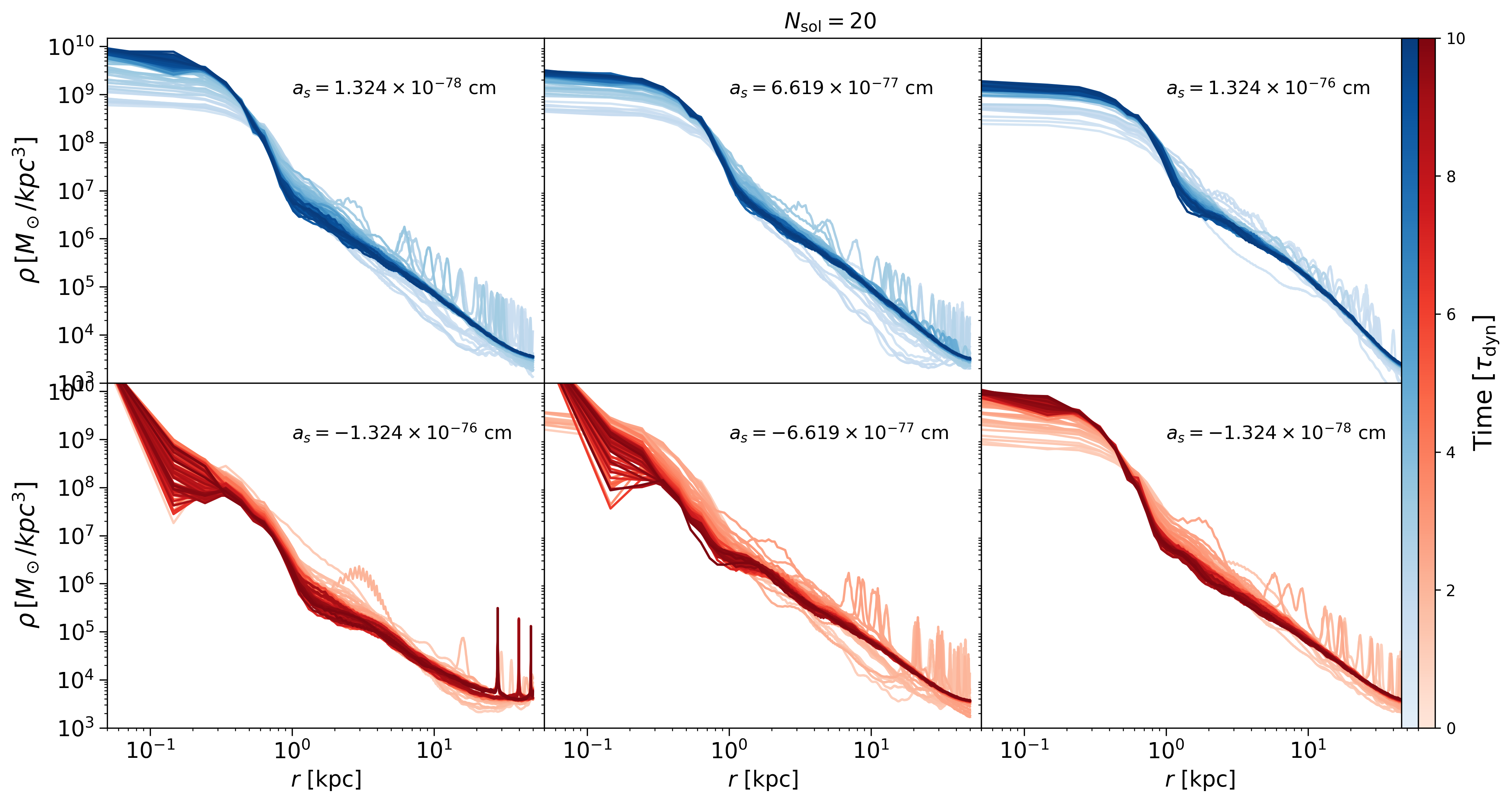}
    \caption{Evolution of the density profiles for repulsive (top panels) and attractive (bottom panels) interactions. Some representative values of $a_s$ were selected from Table \ref{tab: table_params}.
    The profiles are displayed up to the final snapshot in time which corresponds to $10\tau_{\text{dyn}}$ for all simulations. Spikes at large radii trace non-virialized, infalling solitons and interference-dominated streams; they are transient outer-envelope features that damp away as the system relaxes. }
    \label{fig: Evolution_prof}
\end{figure*}

\subsection{Evolution of the density profiles}

In Fig. \ref{fig: Evolution_prof}, we show the evolution of the density profiles over time for some of the parameters considered in Table \ref{tab: table_params}, using $N_{\text{sol}}=20$ for illustration and without loss of generality. For both attractive and repulsive cases, we can observe pronounced oscillations in the density profiles at early times, indicating the merger of solitons in the center. At late times, the central density reaches its maximum value with an extended (flat) core for the repulsive case, making the transition between core and envelope regions more pronounced. This behaviour is in agreement with the results reported in \citet{Guzman2018evm, stallovits2024single}. For the attractive case, we observe a similarly flat core for SI values weaker than $a_s \sim -1\times 10^{-78}$ cm. However, when stronger interactions are considered, the density profile begins to diverge in the center after a few time steps. In fact, the stronger the interactions, the more rapidly the density diverges, in agreement with our expectation that strongly attractive SI leads to different equilibria with the possibility of gravitational collapse of the halo core (such SFDM models provide "core collapse" in the context of galactic halos, see below). 

\begin{figure*}
    \centering
    \includegraphics[width=0.85\linewidth]{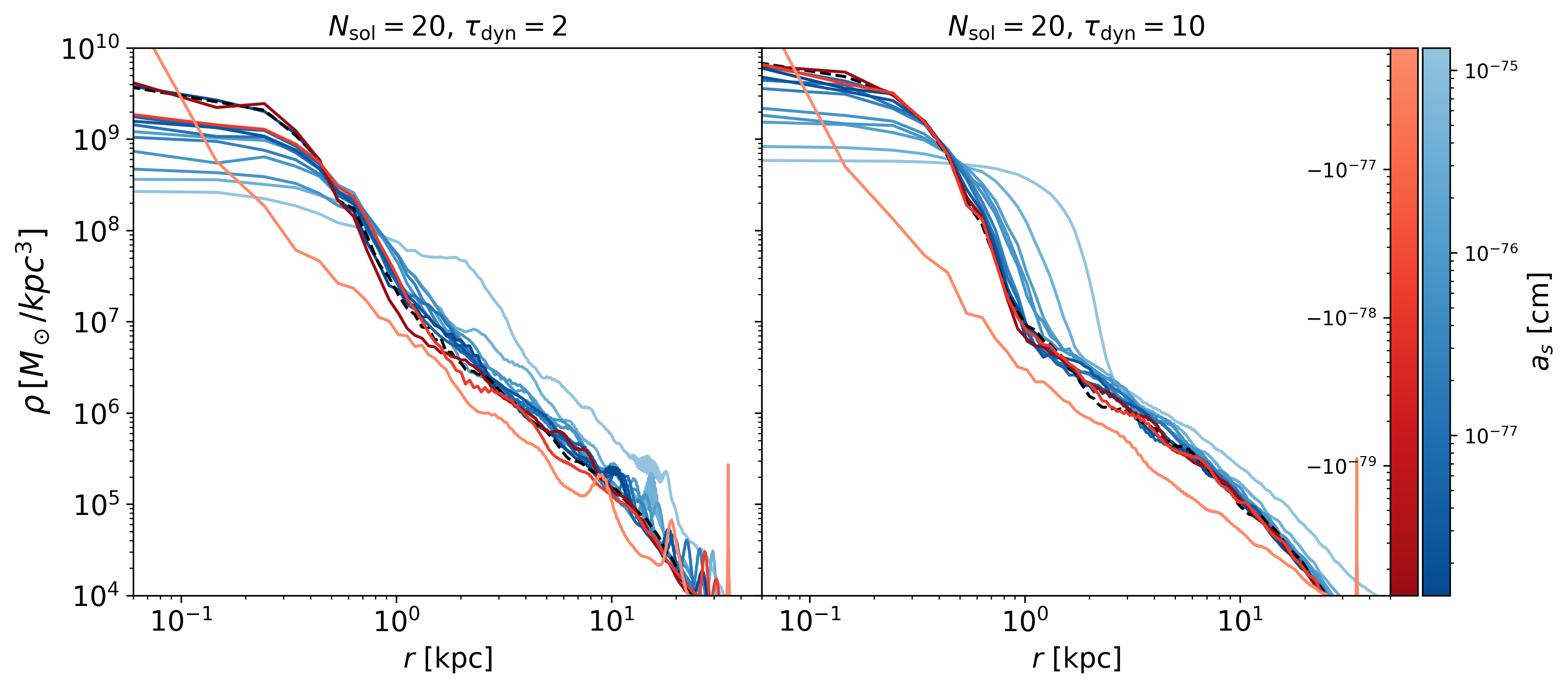}
    \caption{Comparison of the density profile at $2\tau_{\text{dyn}}$ (left panel) and 10$\tau_{\text{dyn}}$ (right panel) for all the values in Table \ref{tab: table_params}. The blue (red) lines show models with repulsive (attractive) SI, while the black dashed curve represents the FDM case ($a_s=0$). At the beginning, the profiles show strong oscillations, particularly at the outskirts, which fade as the halo evolves. The central density increases for the attractive case as the system evolves. The opposite happens for the repulsive scenario, whose central density goes below the FDM regime ($a_s=0$); stronger repulsive SI (i.e. higher positive $a_s$) leads to lower central densities and larger core size. }
    \label{fig: density_times}
\end{figure*}

Furthermore, we compare the density profiles of all parameters at two different moments in the evolution in Fig. \ref{fig: density_times}. 
For the repulsive scenario, the central densities are lower than in the FDM regime ($a_s=0$), leading to more extended cores and a clear core-envelope halo structure, as also seen in \citet{Dawoodbhoy2021beb, 2022Hartman, shapiro2022cosmological, foidl2023halo, stallovits2024single, indjin2025fuzzy}.
On the other hand, for attractive SI, regions near the center exhibit higher densities compared to the FDM case, and this behaviour persists throughout the halo's evolution. Furthermore, the envelopes have somewhat lower densities than in FDM, as the central part accumulates more and more of the halo mass. Eventually, the central density diverges, as previously mentioned.

\begin{figure}
    \centering    \includegraphics[width=0.95\linewidth]{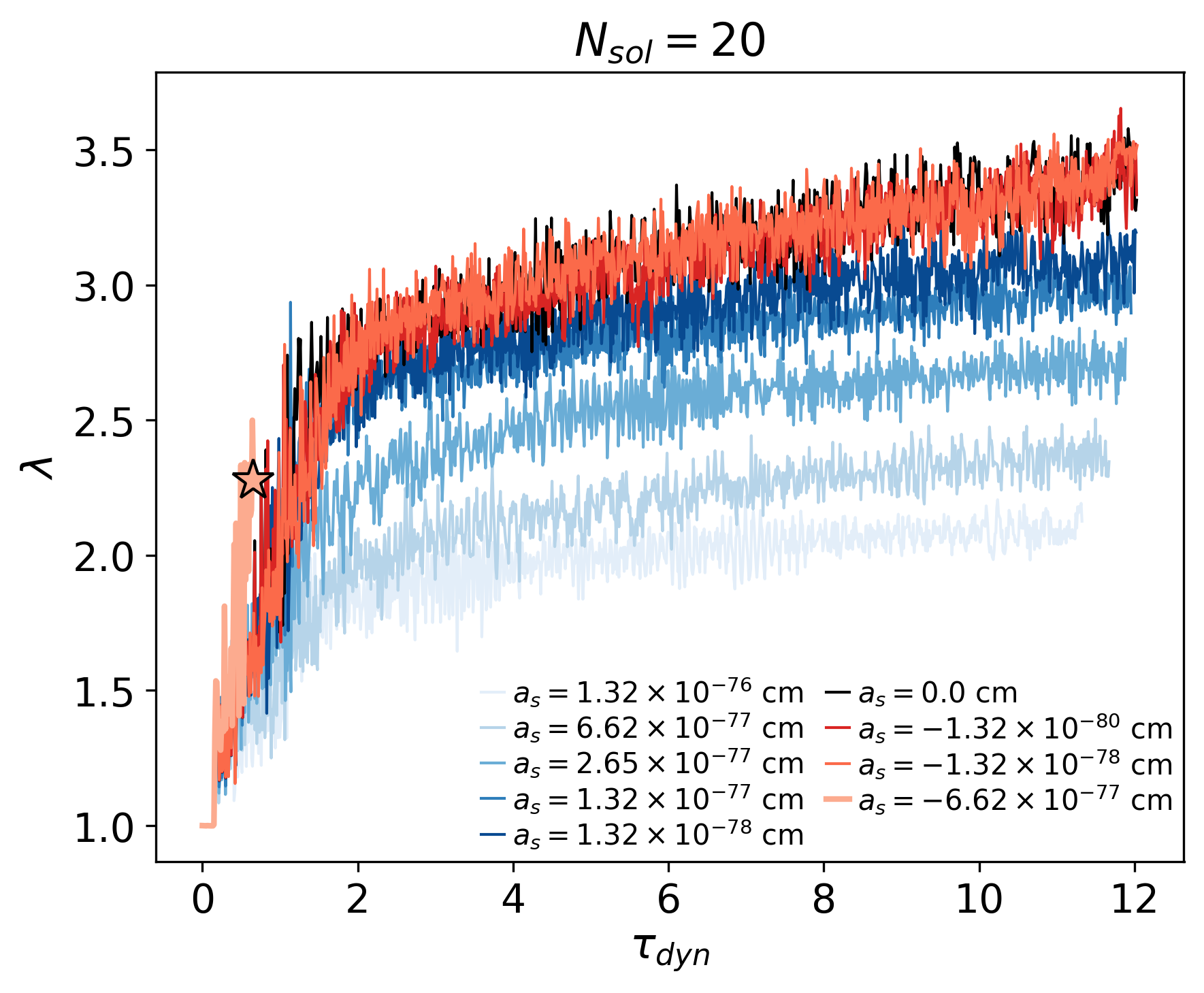}
    \caption{Evolution of the normalized central density given by 
    $\lambda=\left(\frac{\rho(r=0,t)}{\rho(0,0)}\right)^{1/4}$ 
    as a function of time for a subset of representative values in Table \ref{tab: table_params}. The star indicates the moment when $M_{\text{max}}$, according to eq. (\ref{eq: mmax}) is reached.}
    \label{fig: lambda_evolution}
\end{figure}

Fig. \ref{fig: lambda_evolution} shows the evolution of the ratio between the central density and its initial value 
\begin{equation}\label{eq: lambda}
    \lambda=\left(\displaystyle\frac{\rho(r=0,t)}{\rho(0,0)}\right)^{1/4}
\end{equation}
for $N_{\text{sol}}=25$. We note that $\lambda$ is related to the transformations in eq. (\ref{eq: scaling1}). \citet{lopez2025scaling} reported that this quantity follows a power law for the FDM case.
Here, we observe that the power-law exponent depends on the value of SI, again in accordance with our physical understanding: compared to FDM, the density ratio is smaller in the repulsive cases, for the density gradients are much smaller during the evolution. The opposite happens for attractive SI, where the steepening of the density in the central parts enhances the density gradients over time.
This is consistent with the results reported in \citet{chen2021new, painter2024attractive}.

\subsubsection{Critical core mass for gravitational collapse}

It is well known that for SFDM models with repulsive SI, $a_s> 0$, there is always a stable hydrostatic equilibrium configuration for any value of the core mass (or halo mass), in the Newtonian regime, because SI always counterbalances the gravitational force. On the other hand, for models with attractive SI, $a_s<0$, which works on top of the gravitational inward pull, equilibrium can only exist below a certain threshold of the core mass, or of the SI value respectively, even in the Newtonian regime; see
e.g. \cite{chavanis2016collapse},
\begin{equation}\label{eq: mmax}
    M_{\text{max}}=1.012 \frac{\hbar}{\sqrt{G m\abs{a_s}}}.
\end{equation}
Thus, halo cores in this regime can potentially collapse to a central supermassive black hole, during halo evolution, should they accumulate too much mass over time. This formation channel of central, galactic supermassive black holes, or their seeds, as well as astrophysical implications have been studied in 
\citet{Avilez_etal2018, padilla2021core}.
As an illustration, let us consider a few examples of strongly attractive SI, as follows. For $\abs{a_s}>1\times 10^{-77}$ cm, the mass in the core reconstructed from the simulations is close to their corresponding $M_{\text{max}}$. In fact, the density profile diverges before $\lesssim 2 \tau_\text{dyn}$, as can be seen in Fig. \ref{fig: density_times}. In particular, for $a_s=-6.62\times 10^{-77}$ cm, corresponding to $\Lambda = -5 \times 10^6$, the maximum mass corresponds to $M_{\text{max}}=6.04\times 10^8 M_{\odot}$ and it is shown as a star in Fig. \ref{fig: lambda_evolution}, indicating that the collapse occurs around the first $\tau_{\text{dyn}}$, i.e. roughly over one Hubble time. We note that this particular SFDM model, $m=10^{-22}$ eV$/c^2$, $\Lambda = -5 \times 10^6$, with its high critical collapse mass of $M_{\text{max}}=6.04\times 10^8 M_{\odot}$ falls outside the fiducial ("ideal") model range chosen on page 21 in \cite{padilla2021core}, which was designed to produce smaller (seed) black holes, in a range of $10^6-10^7~M_\odot$. Furthermore, our numerical results of the mass-radius relation of the core agree very well with analytical predictions; see Sec.\ref{sec: core} and Fig.\ref{fig: mass-radius}.

\subsubsection{Total energy}

The total energy of the SFDM halo is given by
\begin{equation}\label{eq: total_energy}
    E=K+W+U_{\text{SI}},
\end{equation}
where the first two terms correspond to the kinetic and gravitational potential energy, while the last term is the energy associated with SI. The kinetic energy $K$ is itself composed of the quantum and classical contributions, according to
\begin{align}\label{eq:energy_decomposition}
    \begin{aligned}
        &K = \frac{\rho_0 \hbar^2}{2m^2} \int_V dV\, \nabla\psi^* \cdot \nabla\psi, 
        & \\
        &\phantom{K} = \underbrace{\frac{\rho_0 \hbar^2}{2m^2} \int_V dV\, \nabla\sqrt{\psi\psi^*}}_{K_v} 
        + \underbrace{\frac{\rho_0}{2} \int_V dV\, \psi\psi^* \mathbf{v}^2}_{K_{\rho}}, 
        & \\
        &W = \frac{\rho_0}{2} \int_V dV\, \Phi\, \psi^* \psi, 
        &\\
        &U_{\text{SI}} = \frac{2\pi \hbar^2 \rho_0^2 a_s}{m^3} \int_V dV\, (\psi^* \psi)^2  
        &
    \end{aligned}
\end{align}

$K_v$ accounts for the quantum term associated with density gradients, while the classical term, $K_\rho$, is related to any bulk motion in the halo, and we follow here the notation of \cite{TurbulenceI}. The gravitational potential energy, $W$, represents the binding effect of self-gravity, while the self-interaction energy, $U_{\text{SI}}$, depends linearly on the scattering length $a_s$ (here is no higher-order term included, but see Appendix \ref{app: comparison}).
The sign and magnitude of $a_s$ determine whether this term acts as a repulsive contribution that supports the core, or as an attractive one that enhances gravitational inward pull. Altogether, the balance between these energies governs the stability and dynamical evolution of the halo, see also Sec. 4, Fig. 8.

The stability criterion for the conservation of the energy is discussed in Appendix \ref{app: stability}.



\section{Scaling relations}
\label{sec: core}

In this section, we present scaling relations for the core and the halo at large. Since it is beneficial to compare to a representation of the density profile of halo cores, we use the formula derived from fits to simulated FDM halo density data in \citet{schive2014understanding},
\begin{equation}\label{eq: sol}
     \rho_{\text{core}}(r)=\frac{\rho_c}{\left[1+\alpha \left(\frac{r}{r_c}\right)^2\right]^8},
\end{equation}
where we also fix $\alpha = 0.091$, appropriate for redshift $z=0$. We have defined the center of the final post-merger solitonic halo core as the point with the maximum density. 
In order to determine our scaling relations for the core, we perform a MCMC analysis to find the best values for $r_c$ and $\rho_c$ of eq. \eqref{eq: sol} for all the simulations at all times, with parameters in Table \ref{tab: table_params}. The mass contained in the core is calculated as $M_c=\int \rho_{\text{core}} dV$, while the total mass of the halo is $M=\int\rho dV$, where $\rho=\rho_0\abs{\psi}^2$ corresponds to the 3D mass density. Furthermore, the total energy $E$ is computed numerically in every timestep, using the value of the 3D wave function, through eqs. \eqref{eq: total_energy} and \eqref{eq:energy_decomposition}.

\subsection{Mass-size relation of the core}

As discussed in previous sections, the size and density of the core are influenced by the strength and nature of the SI. Repulsive interactions lead to more dilute cores, while attractive interactions result in more compact and denser cores. In this context, it is particularly insightful to explore how the characteristic core radius $r_c$ and the enclosed mass vary with different values of $a_s$. 

For this comparison, we use the already available analytical expression of the core mass-radius relation from \citet{padilla2021core} This expression was originally derived theoretically in \citet{chavanis2011mass}, and later computed numerically in a hydrodynamic framework in \citet{chavanis2011massII}. It is based upon a Gaussian density distribution for the SFDM solitonic core, 
\begin{equation}\label{eq:Gaussian}
\rho_{\text{core}}(r)=\displaystyle\frac{M_c}{(\pi r_c^2)^{3/2}}e^{-r^2/r_c^2},
\end{equation}
and the relation was found by minimizing the energy in eq. (\ref{eq: total_energy}), leading to
\begin{equation}\label{eq: mc_rc_analytical}
    M_c=3\sqrt{2 \pi}\frac{\hbar^2}{G m^2r_c}\left(\sqrt{2\pi}-\frac{6\hbar^2a_s}{Gm^3r_c^2}\right)^{-1}.
\end{equation}

A detailed discussion of the mass–size relation can be found in the following references \citet{chavanis2020core, chavanis2021jeans,chavanis2023maximum}.

\begin{figure}
    \centering
    \includegraphics[width=\linewidth]{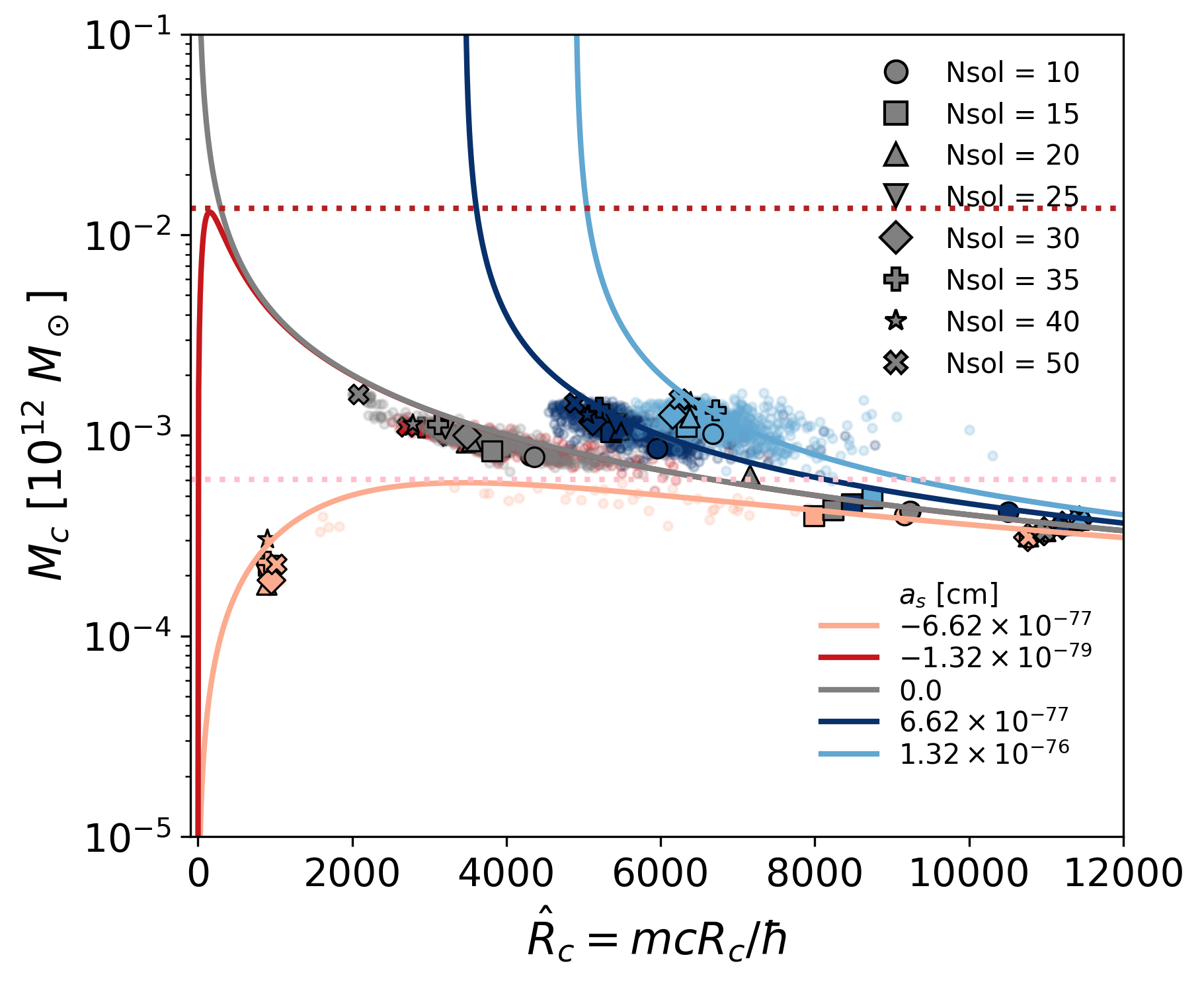}
    \caption{Mass–size relation for SFDM halo cores for the highest values of $a_s$ considered in Table \ref{tab: table_params}, with only a subset shown for clarity. Simulation results are compared with the analytical prediction from eq. (\ref{eq: mc_rc_analytical}), shown as solid curves. More transparent dots correspond to early times, while darker ones represent later times, up to $10\tau_{\text{dyn}}$. The dashed lines represent $M_{\text{max}}$ for the attractive cases.}
    \label{fig: mass-radius}
\end{figure}

We stress that both density distributions, eq. (\ref{eq: sol}) and (\ref{eq:Gaussian}), lead to the same parameter dependences in the $M_c - r_c$ relation, and only the prefactors differ by numbers of order one, a difference which is small in log-plots. 
 In Fig. \ref{fig: mass-radius}, we present both the analytical expression (\ref{eq: mc_rc_analytical}) and the simulation results, showing only the strongest interactions considered in this work for a clearer visualization, for all values $N_{\text{sol}}$ ranging from 10 to 50. In fact, for weaker SI, $\abs{a_s}<1\times 10^{-80}$ cm, the behaviour is very similar to FDM ($a_s = 0$). Additionally, in the figure, the brightest points represent early times, while the darkest circles indicate later times, up to 10$\tau_{\text{dyn}}$. Moreover, the symbols on the right represent the initial soliton in each scenario, marking the start of the simulation, while those on the left correspond to the final state. We recognize the expected result mentioned earlier: As the halo evolves, more mass is accreted onto the central core, which becomes more compact over time. However, this effect is less pronounced for models with repulsive SI, yielding more dilute cores at the same mass, compared to vanishing or negative scattering length. This robustness implies that the mass–radius relation is largely insensitive to the detailed shape of the density profile, with different core configurations leading to the same parametric dependence and differing only by order-unity prefactors. As a result, the size–mass relation of solitonic cores reflects a fundamental equilibrium constraint, rather than the specifics of the halo environment or formation history.  
  FDM models reproduce the well-known scaling law,
$M_c\propto 1/r_c$,
characteristic of self–gravitating solitons without particle interactions \citep{Membrado, guzman2006gravitational}. In contrast, introducing a repulsive SI provides additional pressure that allows more massive configurations at a given core radius, shifting the relation upward with respect to FDM.
Additionally, for a higher value of $N_{\text{sol}}$, the core size decreases while the central density increases (compare to figure 19 in \cite{stallovits2024single}). For the attractive case, the mass in the core is also related to the total number of $N_{\text{sol}}$, but growth is bounded by the maximum mass $M_{\text{max}}$, indicated by the dashed lines. Here, we only show the points for $a_s=-6.62\times 10 ^{-77}$ cm before collapse. In this case, the additional inward pull reduces the maximum stable core mass, shifting the mass-size relation downward. At the same time, the cores become more compact, leading to higher central densities despite their lower total masses (see also Fig. \ref{fig: lambda_evolution}). Overall, the simulation results follow to a good extent the analytical predictions, especially for late times, when the cores are approximately virialized.

\subsection{Core-halo relations}

\subsubsection{FDM core-halo mass relation}
\begin{figure}
    \centering
    \includegraphics[width=\linewidth]{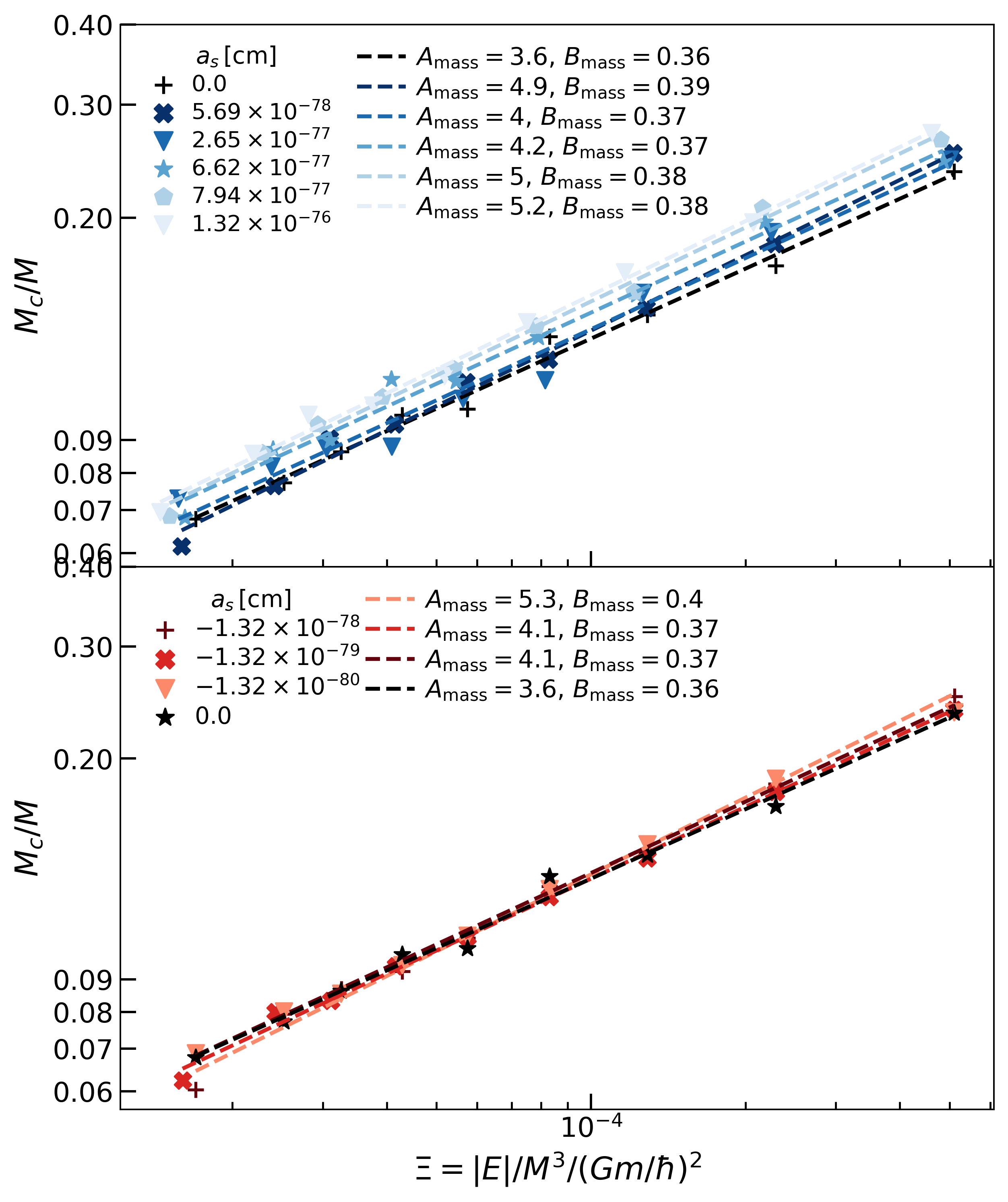}
    \caption{Normalized core mass ($M_c/M$) versus the invariant quantity $\Xi$ for the repulsive (upper panel) and the attractive (lower panel) cases. We can observe a correlation for virialized halos and a slight change in the slope in all cases.}
    \label{fig: core-halo}
\end{figure}

Using numerical simulations for the FDM case, \citet{schive2014understanding} reported an empirical core-halo mass relation of the form $M_c\sim M^{1/3}$ at $z=0$, along with a dependence from the total energy of the system, according to
\begin{equation}\label{eq: schive_energy}
    M_c=\delta\left(\frac{\abs{E}}{M}\right)^{1/2},
\end{equation}
with $\delta\sim 1$. 
However, it has been suggested that this relation may not be fundamental, since it can be directly derived through the scaling parameter $\lambda$ using eq. \eqref{eq: scaling1}. That relation is linked to two strong conditions: the halo is considered a virialized system and the velocity dispersion for the core and the halo satisfy the conditions 
$\sigma_c\sim \sigma_h$ \citep{nori2021scaling} and $\displaystyle\frac{\abs{E_c}}{M_c}\sim \frac{\abs{E}}{M}$ \citep{bar2018galactic} ("velocity-dispersion tracing", \citep{chavanis2019predictive}).

\begin{figure}
    \centering
    \includegraphics[width=\linewidth]{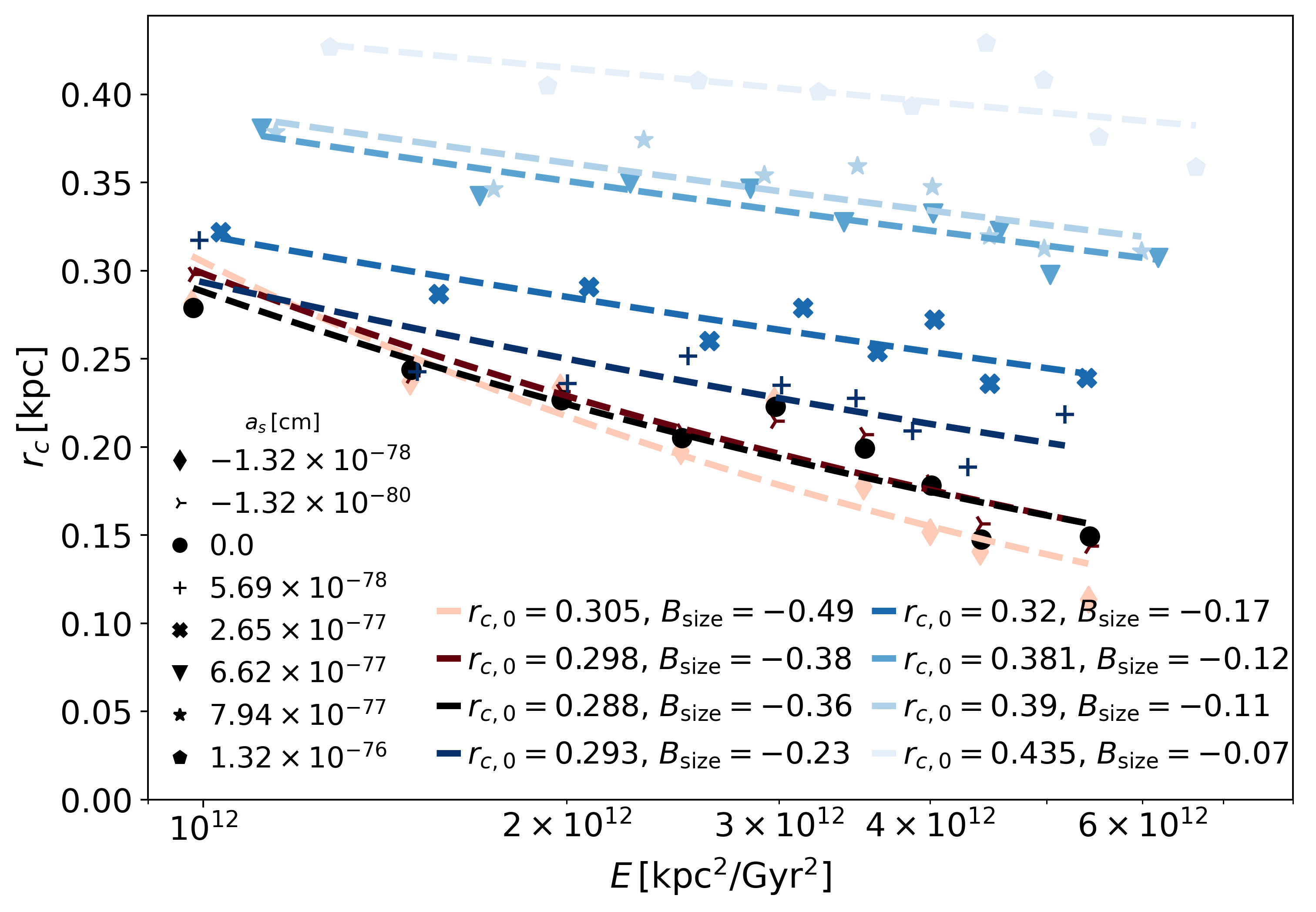}
    \caption{Relation between the scale radius $r_c$ of the core and the total energy of the halo $E$ for some of the attractive and repulsive values from Table \ref{tab: table_params}. Here, all $N_{\text{sol}}$ for each simulation are considered. The best fit for each case is shown as dashed lines, showing a more pronounced dependence for the attractive scenarios.}
    \label{fig: rc_E}
\end{figure}

On the other hand, \citet{mocz2017galaxy} used an invariant quantity under both transformations, eq. (\ref{eq: scaling1}) and eq. (\ref{eq: scaling2}), defined by
\begin{equation} \label{eq:sigmarel}
    \Xi = \left(\displaystyle\frac{\abs{E}}{M^3 (Gm/\hbar^2)}\right)
\end{equation}
to obtain the following (supposedly fundamental) relation for the FDM case
\begin{equation}\label{eq: core-haloMocz}
    \displaystyle\frac{M_c}{M} \sim \Xi^{1/3}.
\end{equation}
As a consequence, in \citet{mocz2017galaxy}, a direct relation is established between the radius and the total energy: $r_c\propto E^{-1/3}$, with $E=K+W$. In a similar way, the most general scenario (compared to \citet{schive2014understanding}) would be described by the relations $\abs{E_c}\sim\abs{E}$ and $M_c\sigma_c^2\sim M\sigma^2$. The previous statements give rise to the relation $M_c\propto M^{5/9}$.  

\begin{figure}
    \centering
    \includegraphics[width=\linewidth]{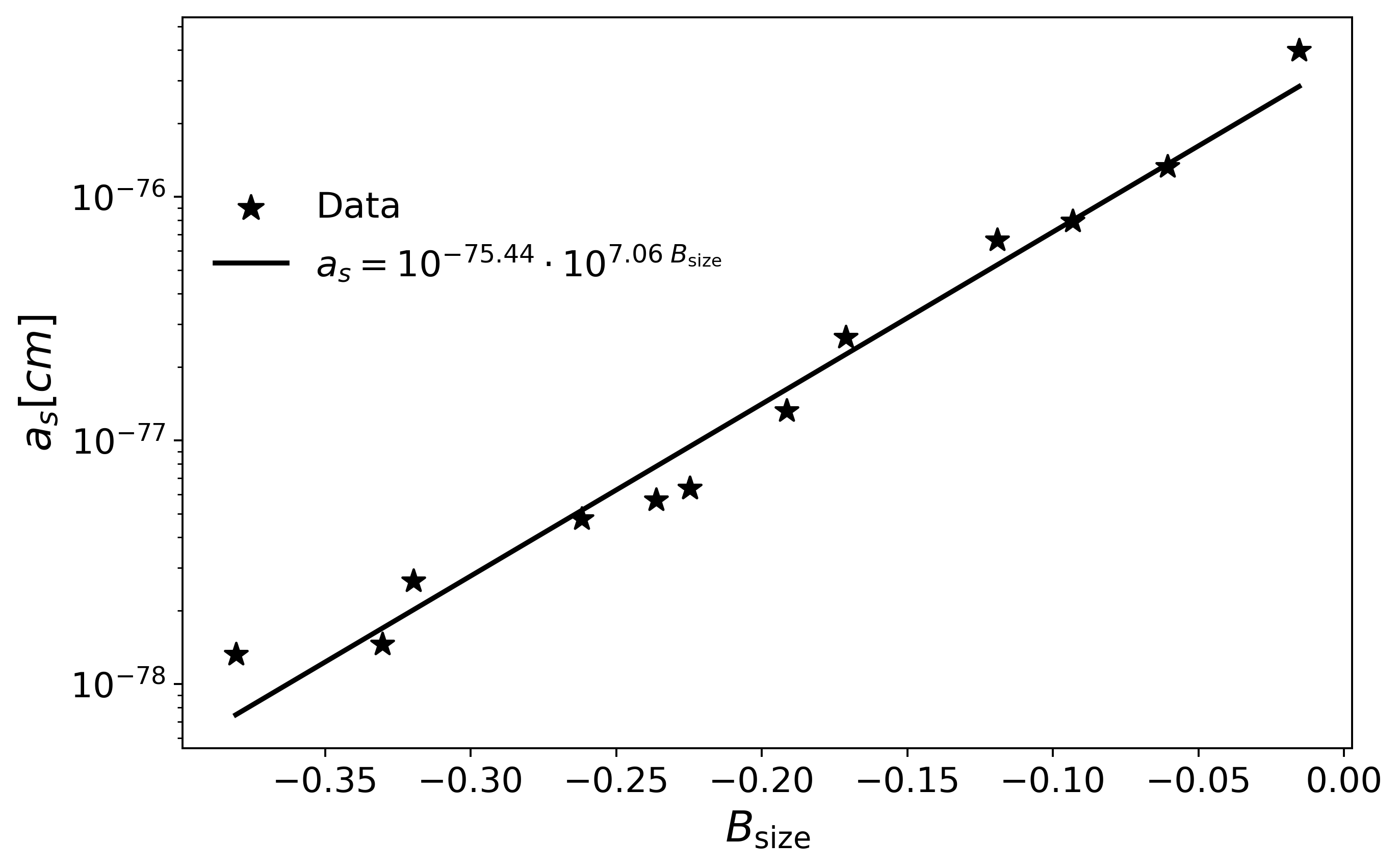}
    \caption{Change of the slope of the $r_c-E$ relation, $B_{\text{size}}$, for different values of the positive scattering length, $a_s$, for model parameters in Table \ref{tab: table_params}. The best fit is also shown as a solid line.}
    \label{fig: m_as}
\end{figure}

 \subsubsection{Self-interacting SFDM core-halo mass relation}

In Fig. \ref{fig: core-halo}, we show the relation between the mass ratio $\displaystyle\frac{M_c}{M}$ and $\Xi$ for different values of the scattering length, as well as the best-fit, using a power law of the form $\displaystyle\frac{M_c}{M}= A_{\text{mass}} \Xi^{B_{\text{mass}}}$. The repulsive (top panel) and attractive (bottom panel) cases are shown separately for a clearer visualization.  
For the FDM case, we find that the cores follow the relation $\displaystyle\frac{M_c}{M}\simeq 3.64\Xi^{0.362}$, which is consistent with eq. (\ref{eq: core-haloMocz}), the \textit{energy-partition} formulation, reported by \citep{mocz2017galaxy}. This regime naturally arises in isolated halos that undergo efficient wave relaxation after multiple soliton mergers, leading to more massive cores of the final halos. In contrast, the well-known relation $M_c\propto M^{1/3}$ derived by \cite{schive2014understanding} in cosmological simulations reflects systems that are continuously fed by accretion and do not fully relax. Thus, rather than being mutually exclusive, the different scaling relations should be regarded as complementary limits within a broader continuum determined by assembly history and relaxation.

Yet, our results show some similarity in the core-halo mass relations, despite different $a_s$, for the multiple soliton merger simulations, within a static background, which we performed. This can be inferred from Fig. \ref{fig: core-halo}, indicating that the mechanism feeding mass to the core is primarily gravitational and therefore universal, once halos are approximately virialized. The SI acts locally in the core, determining its pressure support and radius, independent of the simulation configurations. Thus, we see a mild change in the intercept (normalization), while the slope remains essentially unchanged. 

We stress that we considered the total energy for this computation, including the term due to SI (see eq. \eqref{eq: total_energy}).

\subsubsection{Size-energy relation}

In addition, we study the relations between the size and mass of the core and the total energy of the halo given by eq. (\ref{eq: total_energy}). Figure \ref{fig: rc_E} displays the radius-energy relation at $t=10\tau_{\text{dyn}}$, together with the corresponding power-law fit
$r_c = r_{c,0}\left(\displaystyle\frac{E}{E_0}\right)^{B_{\rm size}}$, where $r_{c,0}=r_c(E_0)$ and $E_0=10^{12}\frac{\text{kpc}^2}{\text{Gyr}^2}$,
 for a subset of representative SI values listed in Table \ref{tab: table_params}. In particular, for $a_s=0$, we have the relation $r_c=0.288\left(\displaystyle\frac{E}{E_0}\right)^{-0.36}$. For repulsive interactions ($a_s>0$), the negative slope is shallower for stronger repulsive SI, and the opposite happens for the attractive case. This is again explained by the fact that, in the repulsive case, solitons are more diluted and extended, showing larger values for $r_c$ at given $E$.

Figure~\ref{fig: m_as} presents the best-fit relation that captures the variation of the slope $B_{\text{size}}$ as a function of the SI parameter, using all positive values of $a_s$ from Table~\ref{tab: table_params}. We find the relation $a_s=10^{-75.44}10^{7.06 B_{\text{size}}}$, which exhibits a clear dependence on the SI strength: for stronger SI, we have larger core sizes and shallower slopes. While SI has little impact on the relative mass distribution within the halo, it plays a significant role in determining the structural properties of the core through their contribution to the internal pressure balance. 

\begin{figure*}
    \centering
    \includegraphics[width=1.0\linewidth]{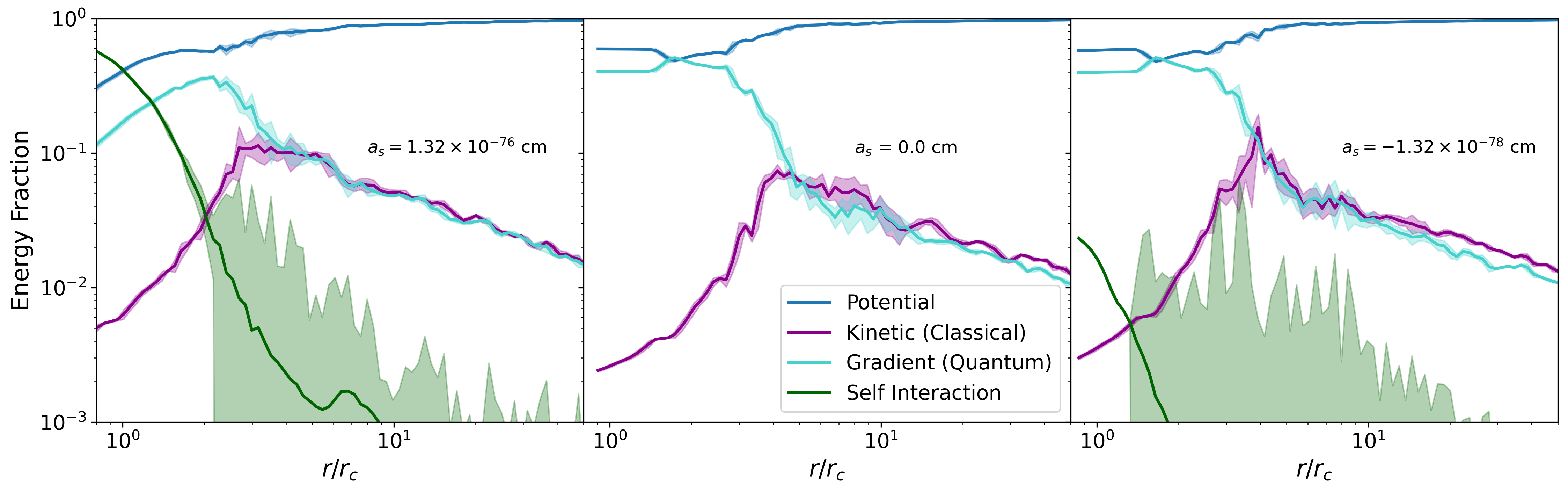}
    \caption{Radially averaged energy fractions for a merger of $N_\text{sol}=20$ solitonic cores, obtained from five equivalent simulations. 
    Each panel corresponds to a different scattering length: $a_s = 1.32\times10^{-76}$ cm (left), $a_s=0$ cm (center), and $a_s=-1.32\times10^{-78}$ cm (right). 
    The colored curves show the relative contribution of the potential (blue), classical kinetic (purple), quantum gradient (cyan), and self–interaction (green) energies, with the shaded regions indicating the $2\sigma$ scatter among simulations. 
    The central panel for $a_s=0$ illustrates the expected ground–state configuration without SI energies. 
    For positive $a_s$ (left panel), the SI energy dominates in the core, while for negative $a_s$ (right panel), the SI energy is very subdominant, thus noisier, for this choice of SI. See main text for more explanations.} 
    \label{fig: contributions}
\end{figure*}

In order to study the dynamics of forces acting on the halo at large, we show in Fig. \ref{fig: contributions} the relative contributions of the different energy components, i.e., potential and kinetic energies, as well as the energies related to SI and quantum term, as described in equations~\eqref{eq:energy_decomposition}. For this energy decomposition, we numerically evaluate the radial profile of each component by spherically averaging the fields. 

We only display the strongest interactions considered in this subsection along with two additional cases: $a_s=1.32\times 10^{-76}$ cm, $a_s=0.0$ cm, and $a_s=-1.32\times 10^{-78}$ cm, all with $N_{\text{sol}}=20$ and at $t=10\tau_{\text{dyn}}$. For each model, we considered five different random seeds as initial position of each soliton.

In all panels, the gravitational term dominates from the outskirts of the core outward and quickly becomes $\geq 90\%$ of the energy budget. The quantum gradient and classical kinetic contributions peak around a few $r_c$ and then decay with radius. In the cases of FDM (middle) and sufficiently weak attractive SI (right panel), we observe that the halo core is dominated by the gravitational potential energy, exemplifying the fact that, for this choice of parameters, halo cores remain in stable gravitational equilibrium. Only in the case of notable repulsive SI (left panel) does the respective SI energy dominate in the halo core, which explains their lower density, larger size and less pronounced variation of $r_c$, as a function of the energy as seen in Fig. \ref{fig: rc_E}. This prevalence of repulsive SI in halo cores has been also found in the SFDM-TF models studied in \cite{shapiro2022cosmological, foidl2023halo} and references therein. In fact, \cite{foidl2023halo} reported a late-time expansion of halo cores, thus indicating some drift away from a well-equilibrated state.

\subsubsection{Variation over time}

\begin{figure}
    \centering
    \includegraphics[width=\linewidth]{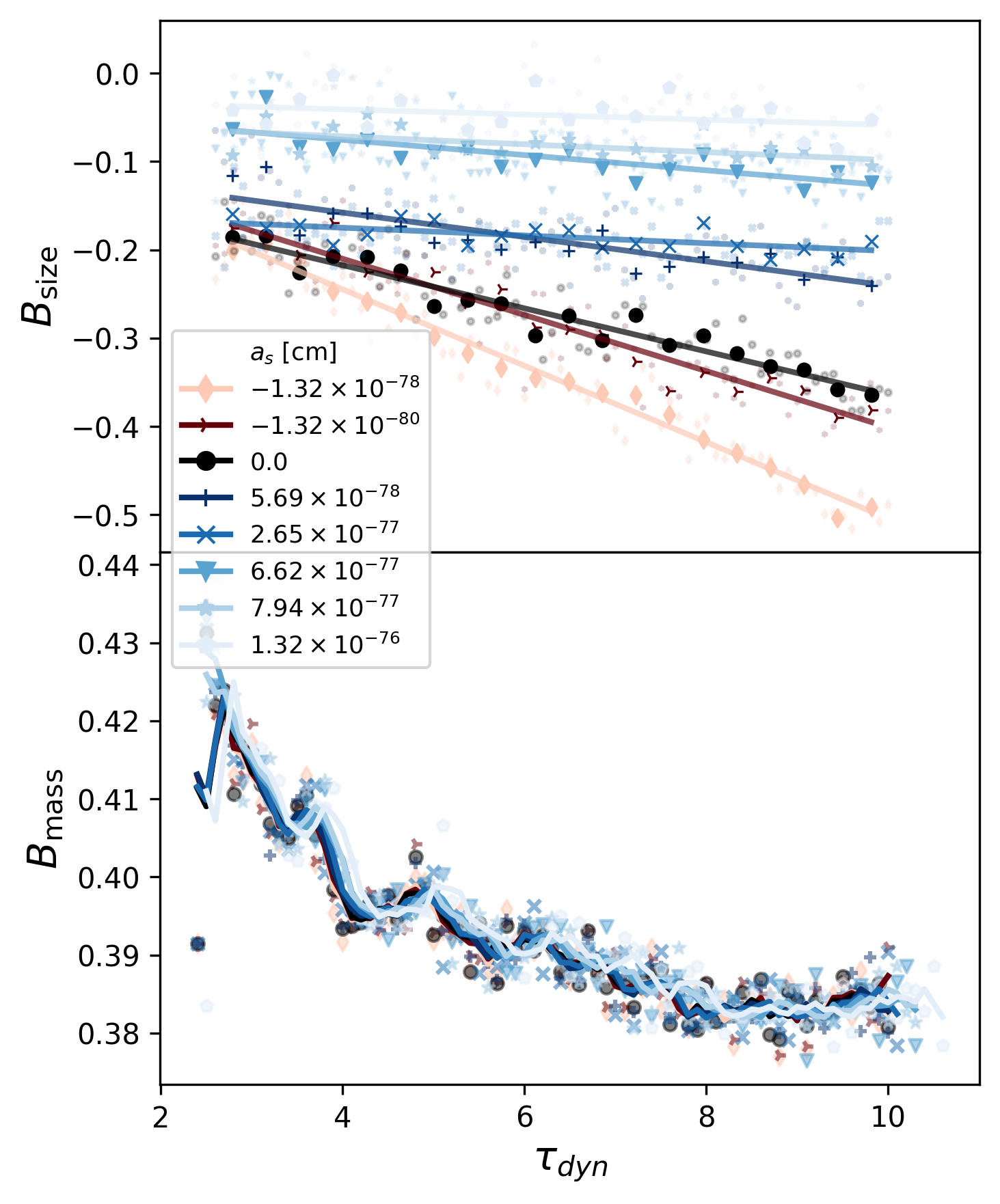}
    \caption{Upper panel: Evolution of the slope $B_{\text{size}}$ of the relation $r_c-E$ over time for different values of the scattering length in Table \ref{tab: table_params}. We observe a decrease, which is more pronounced for the attractive cases.
    Lower panel: Evolution of the slope $B_{\text{mass}}$ of the relation $M_c/M-\Xi$ over time for the same values of $a_s$ as in the upper panel;  again with a decrease over time. }
    \label{fig:B-slope-time}
\end{figure}

The previous relations have been extracted considering $t=10 \tau_{\text{dyn}}$ for all cases, which corresponds to the final snapshot in time of our simulations. However, the core changes in size and mass as the halo evolves. This means that the relations described in the previous sections will also be modified. Therefore, we show the evolution of $B_{\text{size}}$ in the upper panel of Fig. \ref{fig:B-slope-time}. Here, we can see a decrease in the slope over time, being mild for the repulsive cases, but stronger for the attractive and $a_s=0$ cases, especially for the strongest values of SI. 
Additionally, in Fig. \ref{fig:B-slope-time} (lower panel), we observe the change of the slope $B_{\text{mass}}$ over time for different values of SI. The general trend shows a decrease, i.e. the dependence on the global halo parameter $E/M^3$ becomes weaker. This behaviour is similar for all the models shown. 


This analysis indicates that the scaling relations between the core and the halo (whether in terms of size, mass, or density) are closely tied to the evolutionary stage of the halo. As shown in Fig.~\ref{fig: lambda_evolution} and further discussed in the following section, the core gradually accretes mass over time, which in turn shapes the halo at large. In fact, this and previous works show that the final core masses reached in a simulation depend strongly on the initial conditions, and whether mergers happen in a static or expanding background.

\section{Self-similar solutions for core growth} \label{sec: self-similar}

At the end of the last section, we alluded to the evolutionary aspects of core and core-halo scaling relations. In this section, we want to draw a comparison with some previous work, as follows.

\citet{dmitriev2024self} reported an analytical self-similar solution of a Bose star in a thermal bath of gravitationally interacting particles. This solution is valid, provided the parameters drift only slowly over time. In the FDM case of $a_s=0$, the growth of the soliton configuration, thus the core, was determined to follow the relation 
\begin{equation}\label{eq: self_solution}
    \left(1+\left(\frac{M_{BS}}
    {M}\right)^3\frac{1}{E/\gamma M^3}
    \right)\left(1-\frac{M_e}{M}-\frac{M_{BS}}{M}\right)^{-5}\sim \frac{\tau-\tau_i}{\tau_{\star}},
\end{equation}
where $M_{BS}$ is the mass of the boson star, $M_e$ represents the excited bound states in its gravitational field, $\tau_i$ is the initial time, and $\tau_{\star}$ is a timescale to be determined numerically. In that work, the total mass of the system is given by $M=M_{BS}+M_e+M_b$, where $M_b$ is the mass of the particles in the thermal bath with positive energies. We note that the invariant quantity $\sim E/M^3$ appears in this equation, which relates the mass in the core and the halo over time. Additionally, the self-similar solution for models with SI is also mentioned in \citet{dmitriev2024self}. In that case, the authors use an extra function in the cubic term of eq. \eqref{eq: self_solution}, accounting for the SI energy, and determine it numerically.

In our paper, we study the evolution of the mass in the core over time, following an equivalent prescription as in eq. (\ref{eq: self_solution}). However, it is important to point out that we are not considering the same formation process for the halo and its core, as in \citet{dmitriev2024self}. Instead, we use a merger of solitons to create a final halo through the combined effect of gravity, SI and quantum pressure. That means that there are no thermal processes involved and we set $M_b=0$. In this sense, $M_e$ corresponds to the excited gravitationally bound modes that make up the extended envelope or halo. These excited modes are not in thermal equilibrium with an external bath, but rather represent the self-organized structure that arises from the nonlinear gravitational relaxation of the field. Therefore, we use the following equation to describe the growth of the core,
\begin{equation}\label{eq: self_solution_SI}
    \left(1+\left(\frac{M_c}{M}\right)^3\frac{1}{\epsilon^2}\right)\left(1-\frac{M_e}{M}-\frac{M_{c}}{M}\right)^{-5}\sim \frac{\tau-\tau_i}{\tau_{\text{dyn}}},
\end{equation}
where $\epsilon^2=\displaystyle\frac{\abs{E}}{\gamma M^3 (Gm\hbar^2)}=\frac{\Xi}{\gamma}$.

Notice that $\epsilon$ encapsulates the information about the total energy, which in our case also includes the energy due to SI, see eq. \eqref{eq: total_energy}.

\begin{figure}
    \centering
    \includegraphics[width=\linewidth]{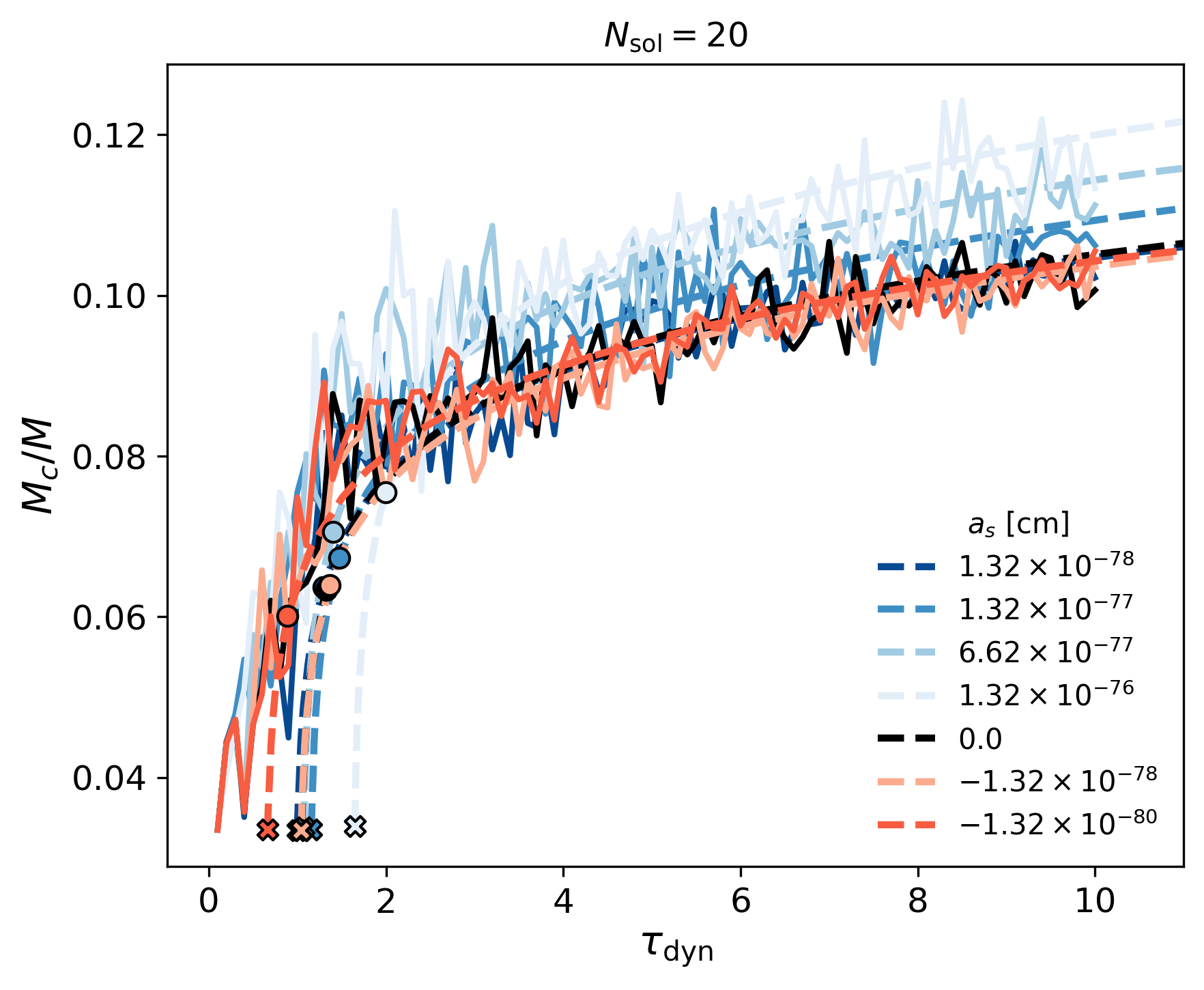}
    \caption{Growth of the relative core mass $M_c/M$ over time for different values of attractive and repulsive self-interactions. The solid lines indicate the simulations, while the dashed lines represent the best fit obtained from eq. (\ref{eq: self_solution_SI}). The first transition from eq. (\ref{eq: self_solution_SI}), $M_c=\epsilon M$, is marked with crosses and the second transition, $M_c=\epsilon^{2/3}M$, is shown with circles. This figure may be compared to Figure S4 in \citet{dmitriev2024self}.}
    \label{fig: self_solutions}
\end{figure}

In our case, $\gamma$ and $M_e$ are parameters to be determined with the fitting procedure. Fig. \ref{fig: self_solutions} shows the evolution of the growth of $M_c/M$ as a function of $\tau_{\text{dyn}}$, for some representative values of the scattering length. This figure may be compared with Figure S4 in the Supplementary Material of \citet{dmitriev2024self}, which shows a similar run. 
We show our fiducial case $N_{\text{sol}}=20$, without loss of generality, with $\epsilon=(
    0.01571,  0.01574, 0.01607, 0.01596, 0.01719, 0.01845, 0.02024 )$
for $a_s= (-1.32\times 10^{-78}, -1.32\times 10^{-80}, 0.0, 1.32\times 10^{-78}, 1.32\times 10^{-77}, 6.62\times 10^{-77}, 1.32\times 10^{-76})$ cm, respectively.
Additionally, notice that the simulations match the fit after $\sim 1 \tau_{\text{dyn}}$, which corresponds to the moment when the solitons have finished merging and the core is formed. 

We observe a clear hierarchy among cases with SI: repulsive interaction leads to a higher relative core mass $M_c/M$, compared to attractive interaction, which are relatively close to FDM ($a_s=0$). This feature is consistent with the findings of \citet{chen2021new}, as repulsive interactions enhance core formation, despite resulting in a lower central density compared to FDM and attractive cases.

Additionally, in \citet{dmitriev2024self} two transitions of boson star growth are mentioned. During the first stage of the evolution, when $E_c \ll E$, the soliton of the resultant configuration is formed and grows until the first transition point, that is, when $\abs{\displaystyle\frac{E_c}{M_c}}=\abs{\displaystyle\frac{E}{M}}$ and the condition $M_c=\epsilon M$ are fulfilled. This relation for the energies corresponds to the condition given by \citet{schive2014understanding}, which, in fact, gives rise to eq. \eqref{eq: schive_energy}. The second transition occurs when $E_c=E$, then $M_c=\epsilon^{2/3}M$. This expression is proportional to eq. \eqref{eq: core-haloMocz} and corresponds to those cores that are usually obtained by merger simulations. In Fig. \ref{fig: self_solutions}, these transitions are shown as a cross for $M_c=\epsilon M$, and as a circle for $M_c=\epsilon^{2/3}M$. Indeed, we observe that the first transition occurs for values of $M_c$ that we cannot reach with the values of the initial solitons we used for our simulations. The second transition occurs right after the merger occurs, once the core is formed. Moreover, the core-halo mass relation of \citet{mocz2017galaxy}
($M_c=\epsilon^{2/3} M$) occurs first for the attractive case for a lower value of $\displaystyle\frac{M_c}{M}$ and last for the strongest repulsive case that we show. The repulsive interaction promotes the formation of the core, but it also prevents a rapid mass accumulation toward the center, leading instead to a more diffuse and massive core, associated with higher values of $\epsilon$.

\section{Conclusions} \label{sec: conclusions}

In this work, we performed multiple SFDM soliton merger simulations within a static background, in order to study various properties of the halos that form in this structure-formation process. We used the pseudo-spectral method to solve the underlying GPP equations of motion, for a large range of values for the self-interaction (SI) of the $s$-wave, $2$-boson scattering. Our study includes the strongly-repulsive Thomas-Fermi regime of SFDM-TF, which has hitherto been modelled using fluid approximations of the GPP equations. The results are in mutual agreement.

We analyzed the resulting halo density profiles, core relations and various core-halo scaling relations, for SFDM models with different values for both attractive and repulsive SI.  
First, we found that repulsive SI gives rise to the formation of more dilute and massive cores with larger size, while the opposite happens for the attractive case, where the high density in the center can even diverge, if the interaction is strong enough. These results are consistent with previous literature, see e.g. \citep{mocz2023cosmological, painter2024attractive, stallovits2024single}.

Additionally, we studied the evolution of the central density in the halo and observed that there is a monotonic behaviour, depending on the value and strength of the SI. Specifically, stronger repulsive SI leads to lower central densities, below those of the FDM model, whereas stronger attractive SI result in higher central densities, exceeding those of FDM.  Again, these results have previously been reported by other authors \citep{chen2021new, dmitriev2024self}. Therefore, our code and procedure have been proven accurate. 

Regarding the mass and size of the cores, our simulations closely follow the analytical prediction presented in \citet{chavanis2011mass} and \citet{ padilla2021core}. This relation was derived assuming a Gaussian density profile and minimizing the total energy, including contributions from SI. We find that the outcome of soliton mergers in our simulations, resulting in overdensities that evolve into halo-like structures, are well described by this theoretical framework. 
Indeed, in the repulsive case, the cores start to accrete mass as they slightly shrink, in agreement with the evolution of the density profile (see Fig. \ref{fig: density_times}). In the attractive case, something similar happens, provided that the critical mass of gravitational collapse is not reached. (When this occurs, there is no equilibrium configuration that can hold the inward pull of gravitational force and SI). For this reason, we only considered values of $a_s$ whose corresponding maximum mass is below this threshold. 

The core-halo mass relation was also explored through the relation between $\displaystyle\frac{M_c}{M}$ and the invariant quantity $\Xi=\displaystyle\frac{\abs{E}}{M^3(Gm/\hbar^2)}$. We fit a power law $\displaystyle\frac{M_c}{M}=A_{\text{mass}}\Xi^{B_{\text{mass}}}$ for different values of $a_s$, finding only a slight variation in the slope \( B_{\text{mass}} \) compared to the FDM case, indicating that once the SI value for \( a_s \) is fixed, the growth of the core mass \( M_c \) relative to the total mass of the halo is primarily governed by gravity. Nevertheless, the relation between the core scale radius $r_c$ and the total energy $\abs{E}$ shows a change in the slope, meaning that the size of the cores (related to the characteristic scale radius) depends on the internal pressure balance among gravitational force, quantum pressure and SI. In fact, we found a relation of the form $a_s=10^{-75.44}10^{7.06B_{\text{size}}}$ for the repulsive models.

This is consistent with the dynamics of the energy components that act on the halo at large. Notably, only in the repulsive case does SI energy dominate in the central region of the halo. Although its contribution can be more than an order of magnitude weaker than that of gravity beyond the core, it still has a noticeable impact on the final density profile and the overall dynamics of the halo.

Additionally, the core-halo mass relation depends on the evolutionary snapshot in time of the halo. The mass in the core increases as the system evolves, until it reaches a transition point where growth is almost constant. On the other hand, $r_c$ changes depending on the contribution of three forces acting on the system. For the repulsive case, we observe that the change in $r_c$ slowly decreases over time. However, in the FDM case, this change is more noticeable, and becomes even more pronounced in the attractive case, since in the latter, SI acts in the same direction as gravity, enhancing the overall potential to collapse. This highlights the differences in the formation and evolution of the cores, when fixing a value of the SI that can give rise to significant differences. These should be tested against observational data in the future. 

Summarising, our results highlight that SI fundamentally alters the balance of forces in SFDM halos, either toward additional support in the repulsive regime, or toward enhanced instability in the attractive one. Consequently, the core–halo mass relation is not universal, but depends on both the SI strength and the evolutionary stage of the system. This indicates that scaling relations derived for free FDM are only limiting cases within a broader family of SFDM models. Beyond reproducing known analytic predictions, our findings suggest that SI provides a natural mechanism to regulate core structure, with implications for the survival of solitonic halo cores, the onset of collapse into supermassive black holes, and the interpretation of astrophysical observations that probe the innermost regions of galaxies. Future work, particularly in cosmological contexts, will be essential to connect these theoretical predictions with observational constraints on the scattering length and particle mass of SFDM.

\section*{Acknowledgements}
JNLS and EMV acknowledge the support by the European Union and the Czech Ministry of Education, Youth and Sports (Project: MSCA Fellowships CZ FZU III -CZ$.02.01.01/00/22\_010/0008598$). This work was supported by the Ministry of Education, Youth and Sports of the Czech Republic through the e-INFRA CZ (ID:90254). TRD acknowledges the support by the Austrian Science Fund FWF through grant nr. P36331-N.

\section*{Data Availability}

The data underlying this article will be shared on reasonable request to the corresponding author.



\bibliographystyle{mnras}
\bibliography{example} 

@article{shapiro2022cosmological,
  title={Cosmological structure formation in scalar field dark matter with repulsive self-interaction: the incredible shrinking Jeans mass},
  author={Shapiro, Paul R and Dawoodbhoy, Taha and Rindler-Daller, Tanja},
  journal={Monthly Notices of the Royal Astronomical Society},
  volume={509},
  number={1},
  pages={145--173},
  year={2022},
  publisher={Oxford University Press}
}

@article{burkert2020fuzzy,
  title={Fuzzy dark matter and dark matter halo cores},
  author={Burkert, Andreas},
  journal={The Astrophysical Journal},
  volume={904},
  number={2},
  pages={161},
  year={2020},
  publisher={IOP Publishing}
}

@article{schive2014understanding,
  title={Understanding the core-halo relation of quantum wave dark matter from 3D simulations},
  author={Schive, Hsi-Yu and Liao, Ming-Hsuan and Woo, Tak-Pong and Wong, Shing-Kwong and Chiueh, Tzihong and Broadhurst, Tom and Hwang, WY Pauchy},
  journal={Physical review letters},
  volume={113},
  number={26},
  pages={261302},
  year={2014},
  publisher={APS}
}

@article{mocz2017galaxy,
  title={Galaxy formation with BECDM--I. Turbulence and relaxation of idealized haloes},
  author={Mocz, Philip and Vogelsberger, Mark and Robles, Victor H and Zavala, Jes{\'u}s and Boylan-Kolchin, Michael and Fialkov, Anastasia and Hernquist, Lars},
  journal={Monthly Notices of the Royal Astronomical Society},
  volume={471},
  number={4},
  pages={4559--4570},
  year={2017},
  publisher={Oxford University Press}
}

@article{schwabe2016simulations,
  title={Simulations of solitonic core mergers in ultralight axion dark matter cosmologies},
  author={Schwabe, Bodo and Niemeyer, Jens C and Engels, Jan F},
  journal={Physical Review D},
  volume={94},
  number={4},
  pages={043513},
  year={2016},
  publisher={APS}
}

@article{du2017core,
  title={Core-halo mass relation of ultralight axion dark matter from merger history},
  author={Du, Xiaolong and Behrens, Christoph and Niemeyer, Jens C and Schwabe, Bodo},
  journal={Physical Review D},
  volume={95},
  number={4},
  pages={043519},
  year={2017},
  publisher={APS}
}

@article{chan2022diversity,
  title={The diversity of core--halo structure in the fuzzy dark matter model},
  author={Chan, Hei Yin Jowett and Ferreira, Elisa GM and May, Simon and Hayashi, Kohei and Chiba, Masashi},
  journal={Monthly Notices of the Royal Astronomical Society},
  volume={511},
  number={1},
  pages={943--952},
  year={2022},
  publisher={Oxford University Press}
}

@article{painter2024attractive,
  title={An attractive model: simulating fuzzy dark matter with attractive self-interactions},
  author={Painter, Connor A and Boylan-Kolchin, Michael and Mocz, Philip and Vogelsberger, Mark},
  journal={Monthly Notices of the Royal Astronomical Society},
  volume={533},
  number={2},
  pages={2454--2472},
  year={2024},
  publisher={Oxford University Press}
}

@article{mocz2023cosmological,
  title={Cosmological structure formation and soliton phase transition in fuzzy dark matter with axion self-interactions},
  author={Mocz, Philip and Fialkov, Anastasia and Vogelsberger, Mark and Boylan-Kolchin, Michael and Chavanis, Pierre-Henri and Amin, Mustafa A and Bose, Sownak and Dome, Tibor and Hernquist, Lars and Lancaster, Lachlan and others},
  journal={Monthly Notices of the Royal Astronomical Society},
  volume={521},
  number={2},
  pages={2608--2615},
  year={2023},
  publisher={Oxford University Press}
}

@article{stallovits2024single,
    author = "Stallovits, Matthias and Rindler-Daller, Tanja",
    title = "{Single and merger soliton dynamics in scalar field dark matter with and without self-interactions}",
    eprint = "2406.07419",
    archivePrefix = "arXiv",
    primaryClass = "astro-ph.CO",
    doi = "10.1103/PhysRevD.111.023046",
    journal = "Phys. Rev. D",
    volume = "111",
    number = "2",
    pages = "023046",
    year = "2025"
}

@article{guzman2004evolution,
  title={Evolution of the Schr{\"o}dinger-Newton system for a self-gravitating scalar field},
  author={Guzman, F Siddhartha and Urena-L{\'o}pez, L Arturo},
  journal={Physical Review D},
  volume={69},
  number={12},
  pages={124033},
  year={2004},
  publisher={APS}
}

@article{chavanis2016collapse,
  title={Collapse of a self-gravitating Bose-Einstein condensate with attractive self-interaction},
  author={Chavanis, Pierre-Henri},
  journal={Physical Review D},
  volume={94},
  number={8},
  pages={083007},
  year={2016},
  publisher={APS}
}

@article{Tidal_O6,
  title = {Tidal disruption of fuzzy dark matter subhalo cores},
  author = {Du, Xiaolong and Schwabe, Bodo and Niemeyer, Jens C. and B\"urger, David},
  journal = {Phys. Rev. D},
  volume = {97},
  issue = {6},
  pages = {063507},
  numpages = {10},
  year = {2018},
  month = {Mar},
  publisher = {American Physical Society},
  doi = {10.1103/PhysRevD.97.063507},
  url = {https://link.aps.org/doi/10.1103/PhysRevD.97.063507}
}

@ARTICLE{Membrado,
       author = {{Membrado}, M. and {Abad}, J. and {Pacheco}, A.~F. and {Saudo}, J.},
        title = "{Newtonian boson spheres}",
      journal = {\prd},
         year = 1989,
        month = oct,
       volume = {40},
       number = {8},
        pages = {2736-2738},
          doi = {10.1103/PhysRevD.40.2736},
       adsurl = {https://ui.adsabs.harvard.edu/abs/1989PhRvD..40.2736M}
}

@ARTICLE{Avilez_etal2018,
       author = {{Avilez}, Ana A. and {Bernal}, Tula and {Padilla}, Luis E. and {Matos}, Tonatiuh},
        title = "{On the possibility that ultra-light boson haloes host and form supermassive black holes}",
      journal = {\mnras},
         year = 2018,
        month = jul,
       volume = {477},
       number = {3},
        pages = {3257-3272},
          doi = {10.1093/mnras/sty572},
archivePrefix = {arXiv},
       eprint = {1704.07314},
 primaryClass = {gr-qc},
       adsurl = {https://ui.adsabs.harvard.edu/abs/2018MNRAS.477.3257A}
}

@article{Edwards_2018,
doi = {10.1088/1475-7516/2018/10/027},
url = {https://dx.doi.org/10.1088/1475-7516/2018/10/027},
year = {2018},
month = {oct},
publisher = {},
volume = {2018},
number = {10},
pages = {027},
author = {Edwards, Faber and Kendall, Emily and Hotchkiss, Shaun and Easther, Richard},
title = {PyUltraLight: a pseudo-spectral solver for ultralight dark matter dynamics},
journal = {Journal of Cosmology and Astroparticle Physics}
}

@article{bar2018galactic,
  title={Galactic rotation curves versus ultralight dark matter: Implications of the soliton-host halo relation},
  author={Bar, Nitsan and Blas, Diego and Blum, Kfir and Sibiryakov, Sergey},
  journal={Physical Review D},
  volume={98},
  number={8},
  pages={083027},
  year={2018},
  publisher={APS}
}

@article{nori2021scaling,
  title={Scaling relations of fuzzy dark matter haloes--I. Individual systems in their cosmological environment},
  author={Nori, Matteo and Baldi, Marco},
  journal={Monthly Notices of the Royal Astronomical Society},
  volume={501},
  number={1},
  pages={1539--1556},
  year={2021},
  publisher={Oxford University Press}
}

@article{desjacques2018impact,
  title={Impact of ultralight axion self-interactions on the large scale structure of the Universe},
  author={Desjacques, Vincent and Kehagias, Alex and Riotto, Antonio},
  journal={Physical Review D},
  volume={97},
  number={2},
  pages={023529},
  year={2018},
  publisher={APS}
}

@article{peccei1977cp,
  title={CP conservation in the presence of pseudoparticles},
  author={Peccei, Roberto D and Quinn, Helen R},
  journal={Physical Review Letters},
  volume={38},
  number={25},
  pages={1440},
  year={1977},
  publisher={APS}
}

@article{weinberg1978new,
  title={A new light boson?},
  author={Weinberg, Steven},
  journal={Physical Review Letters},
  volume={40},
  number={4},
  pages={223},
  year={1978},
  publisher={APS}
}

@article{svrcek2006axions,
  title={Axions in string theory},
  author={Svrcek, Peter and Witten, Edward},
  journal={Journal of High Energy Physics},
  volume={2006},
  number={06},
  pages={051},
  year={2006},
  publisher={IOP Publishing}
}

@article{foidl2023halo,
  title={Halo formation and evolution in scalar field dark matter and cold dark matter: New insights from the fluid approach},
  author={Foidl, Horst and Rindler-Daller, Tanja and Zeilinger, Werner W},
  journal={Physical Review D},
  volume={108},
  number={4},
  pages={043012},
  year={2023},
  publisher={APS}
}

@article{chavanis2011mass,
  title={Mass-radius relation of Newtonian self-gravitating Bose-Einstein condensates<? format?> with short-range interactions. I. Analytical results},
  author={Chavanis, Pierre-Henri},
  journal={Physical Review D—Particles, Fields, Gravitation, and Cosmology},
  volume={84},
  number={4},
  pages={043531},
  year={2011},
  publisher={APS}
}

@article{chavanis2020core,
  title={Core mass-halo mass relation of bosonic and fermionic dark matter halos harboring a supermassive black hole},
  author={Chavanis, Pierre-Henri},
  journal={Physical Review D},
  volume={101},
  number={6},
  pages={063532},
  year={2020},
  publisher={APS}
}

@article{dmitriev2024self,
  title={Self-similar growth of Bose stars},
  author={Dmitriev, AS and Levkov, DG and Panin, AG and Tkachev, II},
  journal={Physical Review Letters},
  volume={132},
  number={9},
  pages={091001},
  year={2024},
  publisher={APS}
}

@article{padilla2021core,
  title={Core-halo mass relation in scalar field dark matter models and its consequences for the formation of supermassive black holes},
  author={Padilla, Luis E and Rindler-Daller, Tanja and Shapiro, Paul R and Matos, Tonatiuh and V{\'a}zquez, J Alberto},
  journal={Physical Review D},
  volume={103},
  number={6},
  pages={063012},
  year={2021},
  publisher={APS}
}

@article{lopez2025scaling,
    author = "L{\'o}pez-S{\'a}nchez, Jessica N. and Munive-Villa, Erick and Skordis, Constantinos and Urban, Federico R.",
    title = "{Scaling relations, dynamical heating, and tidal disruption in spin s ultralight dark matter models}",
    eprint = "2502.03561",
    archivePrefix = "arXiv",
    primaryClass = "astro-ph.CO",
    doi = "10.1093/mnras/staf1616",
    journal = "Mon. Not. Roy. Astron. Soc.",
    volume = "543",
    number = "4",
    pages = "4092--4108",
    year = "2025"
}

@article{chen2021new,
  title={New insights into the formation and growth of boson stars in dark matter halos},
  author={Chen, Jiajun and Du, Xiaolong and Lentz, Erik W and Marsh, David JE and Niemeyer, Jens C},
  journal={Physical Review D},
  volume={104},
  number={8},
  pages={083022},
  year={2021},
  publisher={APS}
}

@article{sipple2025fuzzy,
  title={Fuzzy dark matter constraints from the Hubble Frontier Fields},
  author={Sipple, Jackson and Lidz, Adam and Grin, Daniel and Sun, Guochao},
  journal={Monthly Notices of the Royal Astronomical Society},
  volume={538},
  number={3},
  pages={1830--1842},
  year={2025},
  publisher={Oxford University Press}
}

@phdthesis{winch2024novel,
  title={Novel Cosmological Constraints on Dark Matter},
  author={Winch, Harrison},
  year={2024},
  school={University of Toronto (Canada)}
}

@article{banik2021novel,
  title={Novel constraints on the particle nature of dark matter from stellar streams},
  author={Banik, Nilanjan and Bovy, Jo and Bertone, Gianfranco and Erkal, Denis and De Boer, TJL},
  journal={Journal of Cosmology and Astroparticle Physics},
  volume={2021},
  number={10},
  pages={043},
  year={2021},
  publisher={IOP Publishing}
}

@article{nadler2021constraints,
  title={Constraints on dark matter properties from observations of Milky Way satellite galaxies},
  author={Nadler, EO and Drlica-Wagner, A and Bechtol, K and Mau, S and Wechsler, RH and Gluscevic, V and Boddy, K and Pace, AB and Li, TS and McNanna, M and others},
  journal={Physical review letters},
  volume={126},
  number={9},
  pages={091101},
  year={2021},
  publisher={APS}
}

@article{rogers2021strong,
  title={Strong bound on canonical ultralight axion dark schutz2020subhalomatter from the Lyman-alpha forest},
  author={Rogers, Keir K and Peiris, Hiranya V},
  journal={Physical Review Letters},
  volume={126},
  number={7},
  pages={071302},
  year={2021},
  publisher={APS}
}

@article{schutz2020subhalo,
  title={Subhalo mass function and ultralight bosonic dark matter},
  author={Schutz, Katelin},
  journal={Physical Review D},
  volume={101},
  number={12},
  pages={123026},
  year={2020},
  publisher={APS}
}

@article{TurbulenceI,
    author = {Mocz, Philip and Vogelsberger, Mark and Robles, Victor H. and Zavala, Jesús and Boylan-Kolchin, Michael and Fialkov, Anastasia and Hernquist, Lars},
    title = {Galaxy formation with BECDM – I. Turbulence and relaxation of idealized haloes},
    journal = {Monthly Notices of the Royal Astronomical Society},
    volume = {471},
    number = {4},
    pages = {4559-4570},
    year = {2017},
    month = {07},
    issn = {0035-8711},
    doi = {10.1093/mnras/stx1887},
    url = {https://doi.org/10.1093/mnras/stx1887},
    eprint = {https://academic.oup.com/mnras/article-pdf/471/4/4559/19609125/stx1887.pdf},
}

@article{Rindler-Daller:2011afd,
    author = "Rindler-Daller, Tanja and Shapiro, Paul R.",
    title = "{Angular Momentum and Vortex Formation in Bose-Einstein-Condensed Cold Dark Matter Haloes}",
    eprint = "1106.1256",
    archivePrefix = "arXiv",
    primaryClass = "astro-ph.CO",
    reportNumber = "TCC-035-10",
    doi = "10.1111/j.1365-2966.2012.20588.x",
    journal = "Mon. Not. Roy. Astron. Soc.",
    volume = "422",
    pages = "135--161",
    year = "2012"
}

@article{Dawoodbhoy2021beb,
    author = "Dawoodbhoy, Taha and Shapiro, Paul R. and Rindler-Daller, Tanja",
    title = "{Core-envelope haloes in scalar field dark matter with repulsive self-interaction: fluid dynamics beyond the de Broglie wavelength}",
    eprint = "2104.07043",
    archivePrefix = "arXiv",
    primaryClass = "astro-ph.CO",
    doi = "10.1093/mnras/stab1859",
    journal = "Mon. Not. Roy. Astron. Soc.",
    volume = "506",
    number = "2",
    pages = "2418--2444",
    year = "2021"
}

@article{indjin2025fuzzy,
  title={Fuzzy dark matter haloes with repulsive self-interactions: coherent soliton and halo vortex network with moderate self-coupling},
  author={Indjin, Milos and Keepfer, Nick and Liu, I-Kang and Proukakis, Nick P and Rigopoulos, Gerasimos},
  journal={Monthly Notices of the Royal Astronomical Society},
  volume={545},
  number={3},
  pages={staf2046},
  year={2026},
  publisher={Oxford University Press}
}

@article{galazo2024solitons,
  title={Solitons and halos for self-interacting scalar dark matter},
  author={Galazo Garc{\'\i}a, Raquel and Brax, Philippe and Valageas, Patrick},
  journal={Physical Review D},
  volume={109},
  number={4},
  pages={043516},
  year={2024},
  publisher={APS}
}

@article{navarro1997universal,
  title={A universal density profile from hierarchical clustering},
  author={Navarro, Julio F and Frenk, Carlos S and White, Simon DM},
  journal={The Astrophysical Journal},
  volume={490},
  number={2},
  pages={493},
  year={1997},
  publisher={IOP Publishing}
}

@ARTICLE{2016JiJi,
       author = {{Fan}, JiJi},
        title = "{Ultralight repulsive dark matter and BEC}",
      journal = {Physics of the Dark Universe},
         year = 2016,
        month = dec,
       volume = {14},
        pages = {84-94},
          doi = {10.1016/j.dark.2016.10.005},
       adsurl = {https://ui.adsabs.harvard.edu/abs/2016PDU....14...84F}
}

@ARTICLE{2022Hartman,
       author = {{Hartman}, S.~T.~H. and {Winther}, H.~A. and {Mota}, D.~F.},
        title = "{Cosmological simulations of self-interacting Bose-Einstein condensate dark matter}",
      journal = {\aap},
         year = 2022,
        month = oct,
       volume = {666},
          eid = {A95},
        pages = {A95},
          doi = {10.1051/0004-6361/202243496},
archivePrefix = {arXiv},
       eprint = {2203.03946},
 primaryClass = {astro-ph.CO},
       adsurl = {https://ui.adsabs.harvard.edu/abs/2022A&A...666A..95H}
}

@article{guzman2006gravitational,
  title={Gravitational cooling of self-gravitating Bose condensates},
  author={Guzman, F Siddhartha and Urena-Lopez, L Arturo},
  journal={The Astrophysical Journal},
  volume={645},
  number={2},
  pages={814},
  year={2006},
  publisher={IOP Publishing}
}

@article{glennon2021modifying,
  title={Modifying PyUltraLight to model scalar dark matter with self-interactions},
  author={Glennon, Noah and Prescod-Weinstein, Chanda},
  journal={Physical Review D},
  volume={104},
  number={8},
  pages={083532},
  year={2021},
  publisher={APS}
}

@article{munive2022solving,
  title={Solving the Schr{\"o}dinger-Poisson system using the coordinate adaptive moving mesh method},
  author={Munive-Villa, Erick and L{\'o}pez-S{\'a}nchez, Jessica N and Avilez-L{\'o}pez, Ana A and Guzm{\'a}n, FS},
  journal={Physical Review D},
  volume={105},
  number={8},
  pages={083521},
  year={2022},
  publisher={APS}
}

@article{Guzman2018evm,
    author = "Guzm{\'a}n, F. S. and Avilez, Ana A.",
    title = "{Head-on collision of multistate ultralight BEC dark matter configurations}",
    eprint = "1804.08670",
    archivePrefix = "arXiv",
    primaryClass = "gr-qc",
    doi = "10.1103/PhysRevD.97.116003",
    journal = "Phys. Rev. D",
    volume = "97",
    number = "11",
    pages = "116003",
    year = "2018"
}

@article{zhang2018ultralight,
  title={Ultralight axion dark matter and its impact on dark halo structure in N-body simulations},
  author={Zhang, Jiajun and Tsai, Yue-Lin Sming and Kuo, Jui-Lin and Cheung, Kingman and Chu, Ming-Chung},
  journal={The Astrophysical Journal},
  volume={853},
  number={1},
  pages={51},
  year={2018},
  publisher={IOP Publishing}
}

@article{mocz2018schrodinger,
  title={Schr{\"o}dinger-poisson--vlasov-poisson correspondence},
  author={Mocz, Philip and Lancaster, Lachlan and Fialkov, Anastasia and Becerra, Fernando and Chavanis, Pierre-Henri},
  journal={Physical Review D},
  volume={97},
  number={8},
  pages={083519},
  year={2018},
  publisher={APS}
}

@article{lopez2024estimating,
  title={Estimating the Mass of Galactic Components Using Machine Learning Algorithms},
  author={L{\'o}pez-S{\'a}nchez, Jessica N and Munive-Villa, Erick and Avilez-L{\'o}pez, Ana A and Mart{\'\i}nez-Bravo, Oscar M},
  journal={Universe},
  volume={10},
  number={5},
  pages={220},
  year={2024},
  publisher={MDPI}
}

@article{araya2009cosmology,
  title={Cosmology and cluster halo scaling relations},
  author={Araya-Melo, Pablo A and Van De Weygaert, Rien and Jones, Bernard JT},
  journal={Monthly Notices of the Royal Astronomical Society},
  volume={400},
  number={3},
  pages={1317--1336},
  year={2009},
  publisher={Blackwell Publishing Ltd Oxford, UK}
}

@article{stanek2010massive,
  title={Massive halos in millennium gas simulations: multivariate scaling relations},
  author={Stanek, R and Rasia, Elena and Evrard, AE and Pearce, F and Gazzola, L},
  journal={The Astrophysical Journal},
  volume={715},
  number={2},
  pages={1508},
  year={2010},
  publisher={IOP Publishing}
}

@article{lanzoni2004scaling,
  title={The scaling relations of galaxy clusters and their dark matter halos},
  author={Lanzoni, Barbara and Ciotti, Luca and Cappi, Alberto and Tormen, Giuseppe and Zamorani, G},
  journal={The Astrophysical Journal},
  volume={600},
  number={2},
  pages={640},
  year={2004},
  publisher={IOP Publishing}
}

@article{zagorac2023soliton,
  title={Soliton formation and the core-halo mass relation: An eigenstate perspective},
  author={Zagorac, J Luna and Kendall, Emily and Padmanabhan, Nikhil and Easther, Richard},
  journal={Physical Review D},
  volume={107},
  number={8},
  pages={083513},
  year={2023},
  publisher={APS}
}

@article{matos2024short,
  title={Short review of the main achievements of the scalar field, fuzzy, ultralight, wave, BEC dark matter model},
  author={Matos, Tonatiuh and Ure{\~n}a-L{\'o}pez, Luis A and Lee, Jae-Weon},
  journal={Frontiers in Astronomy and Space Sciences},
  volume={11},
  pages={1347518},
  year={2024},
  publisher={Frontiers Media SA}
}

@article{lee1996galactic,
  title={Galactic halos as boson stars},
  author={Lee, Jae-weon and Koh, In-gyu},
  journal={Physical Review D},
  volume={53},
  number={4},
  pages={2236},
  year={1996},
  publisher={APS}
}

@article{chavanis2025review,
  title={A review of basic results on the Bose--Einstein condensate dark matter model},
  author={Chavanis, Pierre-Henri},
  journal={Frontiers in Astronomy and Space Sciences},
  volume={12},
  pages={1538434},
  year={2025},
  publisher={Frontiers Media SA}
}

@article{chavanis2011massII,
  title={Mass-radius relation of Newtonian self-gravitating Bose-Einstein condensates<? format?> with short-range interactions. II. Numerical results},
  author={Chavanis, Pierre-Henri and Delfini, Luca},
  journal={Physical Review D—Particles, Fields, Gravitation, and Cosmology},
  volume={84},
  number={4},
  pages={043532},
  year={2011},
  publisher={APS}
}

@article{chavanis2021jeans,
  title={Jeans mass-radius relation of self-gravitating Bose-Einstein condensates and typical parameters of the dark matter particle},
  author={Chavanis, Pierre-Henri},
  journal={Physical Review D},
  volume={103},
  number={12},
  pages={123551},
  year={2021},
  publisher={APS}
}

@article{chavanis2023maximum,
  title={Maximum mass of relativistic self-gravitating Bose-Einstein condensates with repulsive or attractive| $\varphi$| 4 self-interaction},
  author={Chavanis, Pierre-Henri},
  journal={Physical Review D},
  volume={107},
  number={10},
  pages={103503},
  year={2023},
  publisher={APS}
}

@article{chavanis2019predictive,
  title={Predictive model of BEC dark matter halos with a solitonic core and an isothermal atmosphere},
  author={Chavanis, Pierre-Henri},
  journal={Physical Review D},
  volume={100},
  number={8},
  pages={083022},
  year={2019},
  publisher={APS}
}

@article{Matos1999et,
    author = "Matos, Tonatiuh and Guzman, Francisco Siddhartha and Urena-Lopez, L. Arturo",
    title = "{Scalar field as dark matter in the universe}",
    eprint = "astro-ph/9908152",
    archivePrefix = "arXiv",
    doi = "10.1088/0264-9381/17/7/309",
    journal = "Class. Quant. Grav.",
    volume = "17",
    pages = "1707--1712",
    year = "2000"
}

@article{peebles2000fluid,
  title={Fluid dark matter},
  author={Peebles, PJE},
  journal={The Astrophysical Journal Letters},
  volume={534},
  number={2},
  pages={L127--L129},
  year={2000}
}

@article{hu2000fuzzy,
  title={Fuzzy cold dark matter: the wave properties of ultralight particles},
  author={Hu, Wayne and Barkana, Rennan and Gruzinov, Andrei},
  journal={Physical Review Letters},
  volume={85},
  number={6},
  pages={1158},
  year={2000},
  publisher={APS}
}

@article{chavanis2019derivation,
  title={Derivation of the core mass-halo mass relation of fermionic and bosonic dark matter halos from an effective thermodynamical model},
  author={Chavanis, Pierre-Henri},
  journal={Physical Review D},
  volume={100},
  number={12},
  pages={123506},
  year={2019},
  publisher={APS}
}



\appendix
\section{Higher-order term}\label{app: comparison}
\begin{figure}
    \centering
    \includegraphics[width=\linewidth]{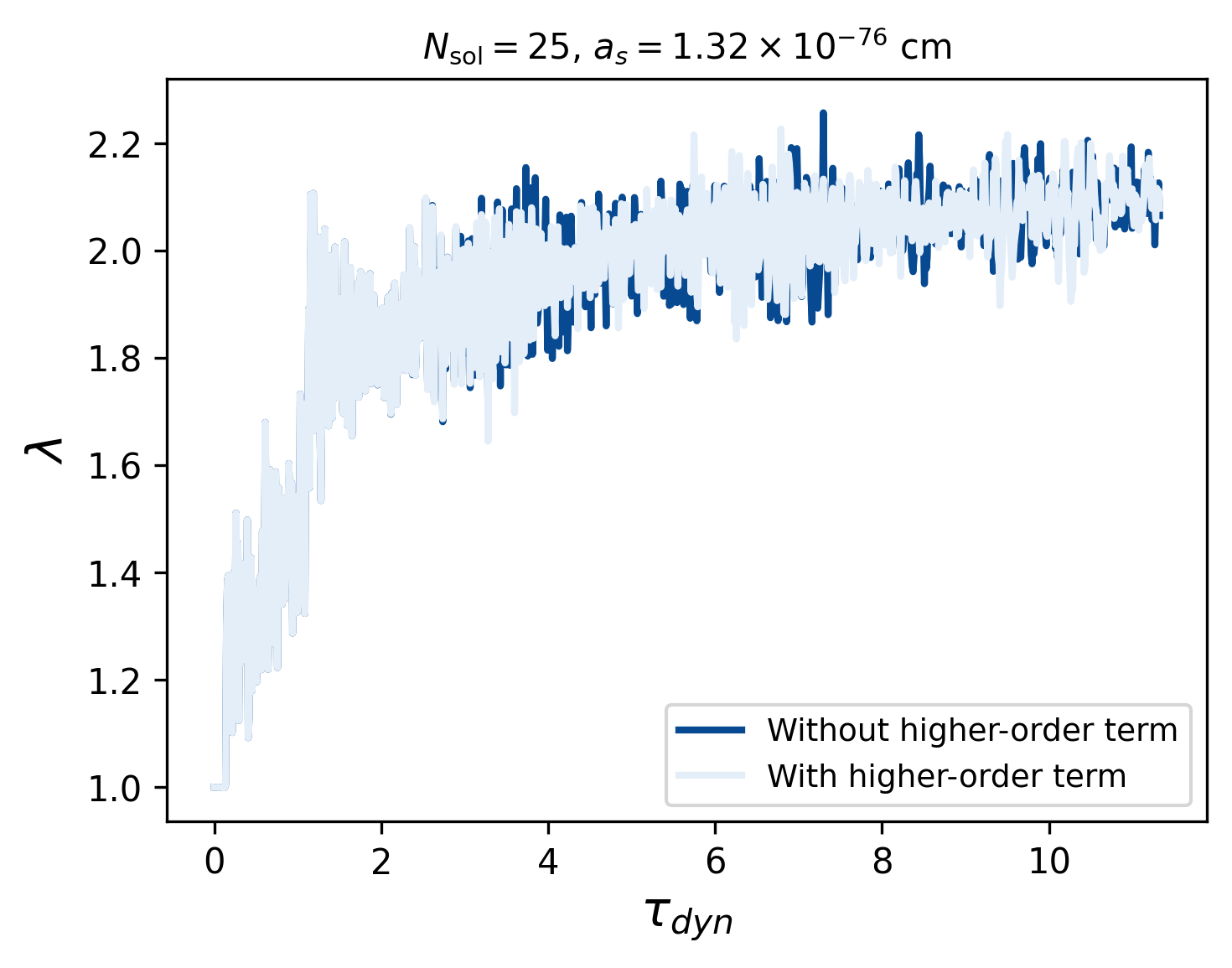}
    \caption{Comparison of the evolution of the density ratio given by eq. (\ref{eq: lambda}) with and without the quintic term ($\sim \psi^5$) in the Gross-Pitaevskii equation discussed in section \ref{sec: GPP}. The simulations were performed for $a_s=1.32\times 10^{-76}$ cm with $N_{\text{sol}}=25$. The curves look similar, exemplifying that the higher-order term can be neglected. }
    \label{fig: order_comparison}
\end{figure}

Figure~\ref{fig: order_comparison} shows the time evolution of the density ratios, defined in eq. (\ref{eq: lambda}), to assess the impact of the higher-order SI term. For clarity, only the strongest repulsive case is presented. The evolution of the density ratios is nearly identical in both cases, indicating that, for the interaction strengths considered here, the inclusion of the higher-order term has a negligible effect.

\begin{figure}
    \centering
    \includegraphics[width=\linewidth]{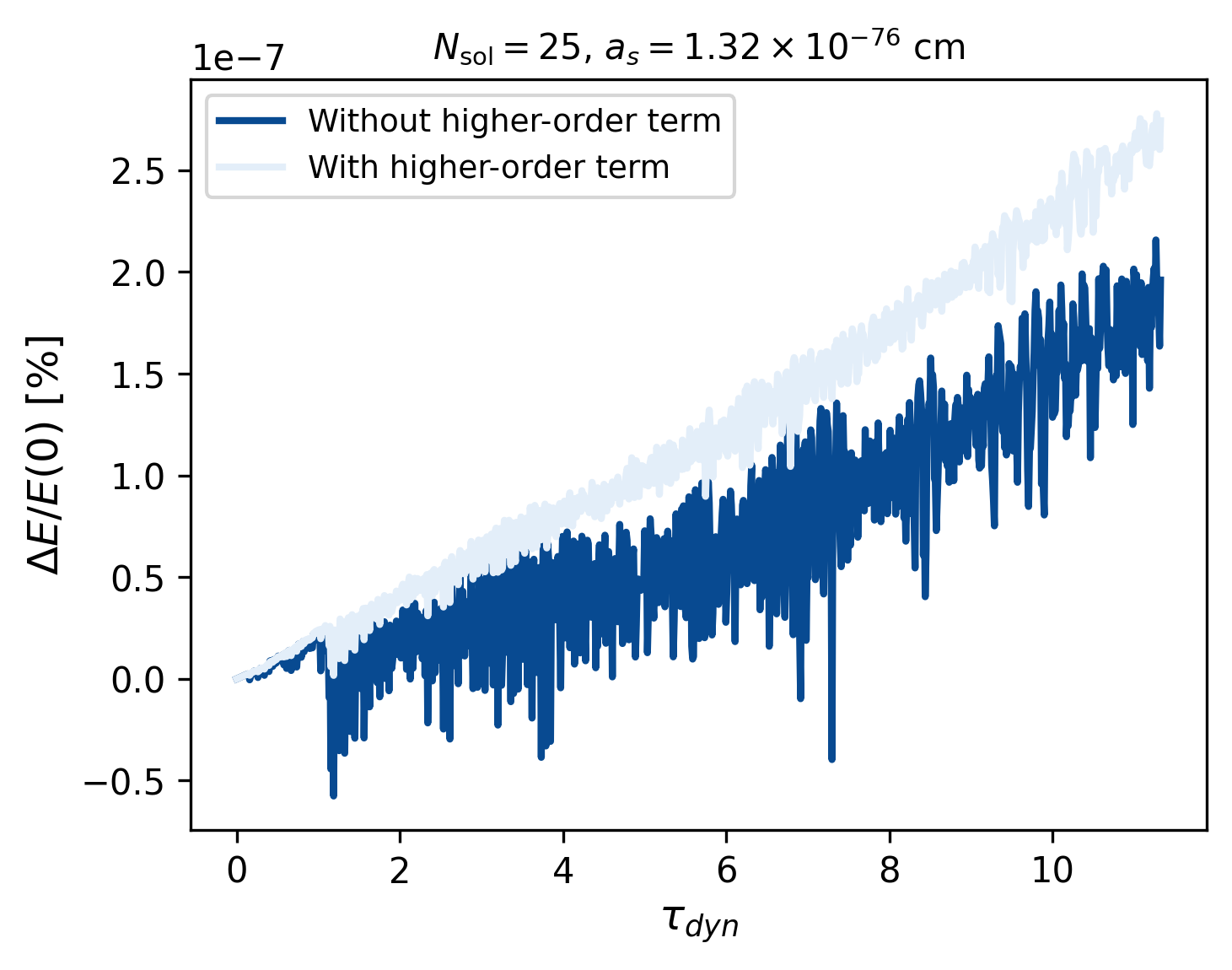}
    \caption{Comparison of the evolution of the energy ratio for both cases, with and without the higher-order term, for the same simulations as in Fig. \ref{fig: order_comparison}. Though there is a slight difference, here the same conclusion applies as in the previous figure.}
    \label{fig: order_energy}
\end{figure}

Similarly, Figure~\ref{fig: order_energy} presents energy conservation through the coefficient $\Delta E / E(0)$ as a function of time. Although small differences are visible, the deviations remain within a very low order of magnitude, further confirming that the effect of the higher-order SI term can be safely neglected.

\section{Stability tests for the solitons}\label{app: stability}
\begin{figure}
    \centering
    \includegraphics[width=\linewidth]{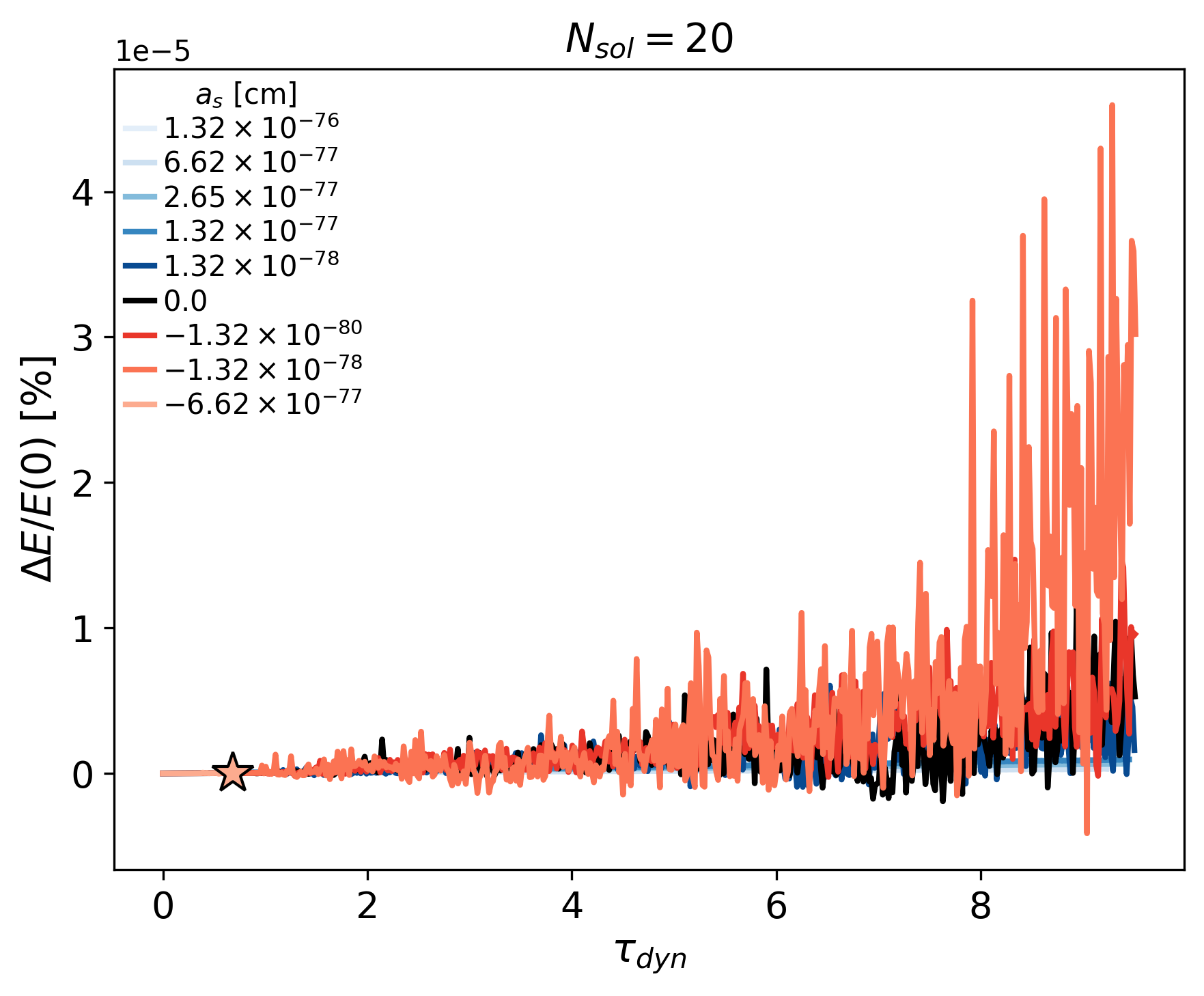}
    \caption{Relative change of the energy with respect to its initial value over time for different values of $a_s$. The star indicates the moment of collapse, once the maximum value of the mass in eq. (\ref{eq: mmax}) is reached for the strongest attractive case considered in this work.   }
    \label{fig: energy_stability}
\end{figure}

To assess the stability of the simulations, we monitored the relative variation of the total energy with respect to its value at the initial time $\displaystyle\frac{\Delta E}{E(0)}$ as a function of the dynamical time $\tau_{\text{dyn}}$ for different values of the SI parameter $a_s$. The results are shown in Fig. \ref{fig: energy_stability}. The deviations remain small, typically below a few $10^{-5}$, which demonstrates that the numerical implementation conserves energy with high accuracy. The noise increases with the magnitude of the SI, especially in the attractive case, however, this is expected as a result of accumulated numerical noise. Importantly, no qualitative differences are observed among the different interaction strengths, which confirms that the calculational procedure remains stable and reliable within the explored parameter range. Note that the case with the strongest attraction remains stable until $\sim 1 \tau_{\text{dyn}}$, just before collapse. This moment is indicated by a star. Beyond this point, both the energy and the density diverge, as shown in Fig. \ref{fig: lambda_evolution}.

\bsp	
\label{lastpage}
\end{document}